# Mapping fat-water separated $R_1$, $R_2^*$, and proton density fat fraction with the multi-echo MP2RAGE sequence

Jorge Campos Pazmiño[1,2], Marc-Antoine Fortin[1], Véronique Fortier[1,3,4], Evan McNabb[1], Renée-Claude Bider[1], André van der Kouwe[5,6], and Ives R. Levesque[1,2,4,7]

[1]*Medical Physics Unit, McGill University, Montréal, QC, Canada*

[2]*Department of Physics, McGill University, Montréal, QC, Canada*

[3]*Department of Medical Imaging, McGill University Health Centre and Department of Radiology, McGill University, Montréal, QC, Canada*

[4]*Gerald Bronfman Department of Oncology, McGill University, Montréal, QC, Canada*

[5]*Athinoula A. Martinos Center for Biomedical Imaging, Department of Radiology, Massachusetts General Hospital, Boston, MA, United States*

[6]*Harvard Medical School, Boston, MA, United States*

[7]*Research Institute of the McGill University Health Centre, Montréal, QC, Canada*

July 8, 2026

* Correspondence to:

Ives R. Levesque, PhD

Medical Physics Unit, McGill University

Cedars Cancer Centre - Glen Site, DS1.9326

1001 Boul. Décarie

Montréal, QC H4A 3J1

ives.levesque@mcgill.ca

514-934-1934 ext. 48105

## Abstract

**Purpose:** To develop a technique for joint measurement of fat and water-specific longitudinal relaxation rates ($R_{1f}$ and $R_{1w}$), effective transverse relaxation rate ($R_2^*$), and proton density fat fraction (PDFF) combining the Multi-Echo Magnetization Prepared Two Rapid Acquisition of Gradient Echoes (ME-MP2RAGE) sequence and fat-water separation.

**Theory and Methods:** $R_{1f}$ and $R_{1w}$ were calculated with fat-specific and water-specific MP2RAGE signals. $R_2^*$ and PDFF maps were obtained from fat-water separation applied to the second RAGE block. Sequence parameters optimization was performed via Cramér-Rao lower bounds theory, and we designed four protocols with different combinations of number of echoes and readout gradient schemes (I: 3 echoes unipolar, II: 6 echoes unipolar, III: 6 echoes bipolar, and IV: 10 echoes bipolar). We tested and validated these protocols with numerical simulations, phantom, and in vivo experiments. In phantoms, we compared ME-MP2RAGE measurements with inversion recovery spin-echo (IR-SE) global $R_1$ and 3D Fast Low Angle Shot (3D FLASH) $R_2^*$ and PDFF. In vivo, we scanned the lower leg and neck of a healthy volunteer.

**Results:** Numerical simulations showed accurate quantification of relaxation rates with mean relative bias < 3% and PDFF with mean bias < 0.003 using protocol ME-MP2RAGE IV (10 echoes bipolar). Phantom experiments showed excellent agreement with IR-SE and 3D FLASH measurements. In vivo, measurements in the lower leg and neck were consistent with literature values.

**Conclusion:** We proposed an accurate method for simultaneous quantification of $R_{1f}$, $R_{1w}$, $R_2^*$, and PDFF from a single acquisition with the ME-MP2RAGE sequence.

## 1 Introduction

Simultaneous measurement of fat and water-specific longitudinal relaxation rates ($R_{1f}$ and $R_{1w}$), effective transverse relaxation rate ($R_2^*$), and proton density fat fraction (PDFF) enables MRI-based applications like MR-oximetry [1–6], study of metabolic processes in adipose tissue [7–9], diagnosis of liver steatosis [10], and assessment of skeletal muscle composition [11–13]. $R_1$ captures changes in tissue oxygenation [1–6], temperature variations in adipose tissue due to metabolic activity [14], and fibrosis in liver or muscle [11,15]. Modeling fat and water proton pools separately increases sensitivity to oxygenation changes via $R_{1f}$ [4] and eliminates fat-induced bias when using $R_{1w}$ to characterize tissue structure in liver or muscle [11,15]. $R_2^*$ reflects blood oxygenation [3,16] and quantifies hepatic iron [17,18]. Finally, PDFF facilitates differentiation between white and brown adipose tissue [19] and allows assessment of fat infiltration in liver [20] and muscle [21].

Inversion recovery (IR) is considered as the “gold standard” technique for global $R_1$ mapping [22–24]. However, its conventional 2D implementation requires long repetition times (TRs) and acquires data at only one effective inversion time (TI) per TR, resulting in prolonged acquisitions [25]. Variations of IR approaches like magnetization-prepared two rapid acquisition of gradient echoes (MP2RAGE) enable fast $R_1$ mapping. The MP2RAGE sequence consists of an inversion pulse followed by two gradient echo readout blocks (RAGE blocks), each generating images at different TIs [26]. MP2RAGE-based $R_1$ mapping is robust against confounding effects due to main field ($B_0$) inhomogeneity, and receive and first-order transmit $B_1$ ($B_1^+$) field inhomogeneity [26]. Many research groups have experimented with adding multiple echoes (per excitation) in both RAGE blocks with monopolar [27] and bipolar [28,29] readout schemes to map $R_2^*$ and magnetic susceptibility. The use of bipolar readout gradient pulses in the MP2RAGE sequence increases the data acquisition efficiency and reduces echo times [29].

Joint measurement of $R_{1f}$, $R_{1w}$, $R_2^*$, and PDFF has been proposed with a variety of approaches: magnetic resonance fingerprinting (MRF) [30–33], magnetic resonance multitasking [34], and combining $R_1$ mapping techniques with chemical shift encode (CSE) fat-water separation [5,22,35,36]. While fast, MRF or MR multitasking typically require complex sequence design and computationally demanding reconstruction [35,37,38]. Combining conventional mapping techniques with CSE fat-water separation represents a more accessible approach, reducing the challenges that hinder the clinical adoption of more advanced methods. Furthermore, this strategy may help establish a standardized reference approach for fat- and water-separated $R_1$ quantification, addressing an important gap in the current literature. However, current techniques are sensitive to $B_1^+$ inhomogeneities [39,40] or treat $R_{1f}$, $R_2^*$, and PDFF as nuisance parameters that confound $R_{1w}$ measurements [22,41–43].

The objective of this work was to combine the multi-echo MP2RAGE (ME-MP2RAGE) sequence with a CSE fat-water separation method to enable simultaneous mapping of $R_{1f}$, $R_{1w}$, $R_2^*$, and PDFF. In this work, PDFF quantification includes corrections for $R_1$-bias [44,45] and noise-bias [44]. We also evaluated the effect of bipolar readout gradient pulses (in the multi-echo portion of both RAGE blocks [29]) on the accuracy and precision of the estimates. This study first presents the theoretical background for multiparametric mapping using the ME-MP2RAGE sequence, followed by an approach for sequence parameter optimization, and finally, testing and validation of the technique with numerical simulations and experiments in a phantom and in vivo.

## 2 Theory

We extend the ME-MP2RAGE signal model to account for fat and water signal compartments according to Eq. 1 [46]. In this equation, $W_r$ and $F_r$ are the complex water and fat signal components that are independent of echo time in each RAGE block [26,47], with the subscript $r = 1, 2$ representing the number of the RAGE block. The subscript $k$ represents the number of the echo (with echo time $TE_k$ and K total echoes). The expression $\sum_{q=1}^{Q} \alpha_q e^{i2\pi f_q TE_k}$ corresponds to the chemical shift effect of a multi-resonance fat spectrum with $Q$ distinct resonances, where $\alpha_q$ and $f_q$ are their relative amplitude and chemical shift. The term $\psi$ represents off-resonance frequency effects due to $B_0$ inhomogeneities. The signal model assumes a common $R_2^*$ term for fat and water relaxation, which prioritizes the stability and noise performance of the fat-water separation technique [48–50]. $\theta = \phi - i\varepsilon$ is a complex term containing the phase error $\phi$ and amplitude modulation $\varepsilon$ introduced by the bipolar readout for the multi-echo portion of the sequence [46,51,52]. For unipolar readout gradients, it is assumed that $\theta = 0$ [46]. In this work, chemical-shift and field-inhomogeneity induced image misregistration were neglected under the assumption that high receiver bandwidth (rBW) is used for the data acquisition [51].

$$\mathrm{Signal_r(TE_k)} = \left( \mathrm{W_r} + \mathrm{F_r} \sum_{q=1}^{Q} \alpha_q e^{i2\pi f_q TE_k} \right) e^{i2\pi\psi TE_k} e^{-R_2^* TE_k} e^{(-1)^k i\theta} \qquad 1$$

Fat-water separation can be applied to each multi-echo RAGE block independently, generating complex fat and water signals for the RAGE1 and RAGE2 blocks: $F_1$, $F_2$, $W_1$, and $W_2$, respectively. Following fat-water separation, the fat and water signals can be combined to generate fat- and water-specific MP2RAGE signals according to equations [26]:

$$\mathrm{MP2RAGE_F} = \mathrm{Real}\left( \frac{F_1^* F_2}{|F_1|^2 + |F_2|^2} \right) \qquad 2$$

$$\mathrm{MP2RAGE_W} = \mathrm{Real}\left( \frac{W_1^* W_2}{|W_1|^2 + |W_2|^2} \right) \qquad 3$$

Fat- and water-specific $T_1$ values can be obtained via linear interpolation with a single lookup table (LUT) linking $T_1$ to the MPRAGE signal calculated in Eq. 2 and 3 [26]. The LUT is generated by computing the MP2RAGE signal for a discrete set of input $T_1$ values, spanning the expected ranges of both fat and water $T_1$. The MP2RAGE signal is calculated first using the signal equations for RAGE1 and RAGE2 (Supplementary Information, Eq. A.1-A.3) with the known sequence parameters (listed in Table 1 and B.1) and plugging the outputs of these into the MP2RAGE signal equation (Eq. 2 and 3). In this work, $R_1$ maps were derived from $T_1$ maps by calculating their voxel-wise reciprocal value.

In principle, PDFF and $R_2^*$ maps can be obtained from the fat-water separation results obtained from either RAGE block. The RAGE2 block was preferred in this work due to the greater PD-weighting and SNR compared to RAGE1. In this work, the signal fat fraction (SFF) was calculated as $|F_2|/(|W_2| + |F_2|)$[53]. The SFF maps were transformed into PDFF maps through corrections to eliminate noise- and $R_1$-bias similar to a previous approach [44], as detailed in Supplementary Information. A visual summary of the proposed technique for multiparametric mapping with the ME-MP2RAGE sequence is presented in Fig. 1.

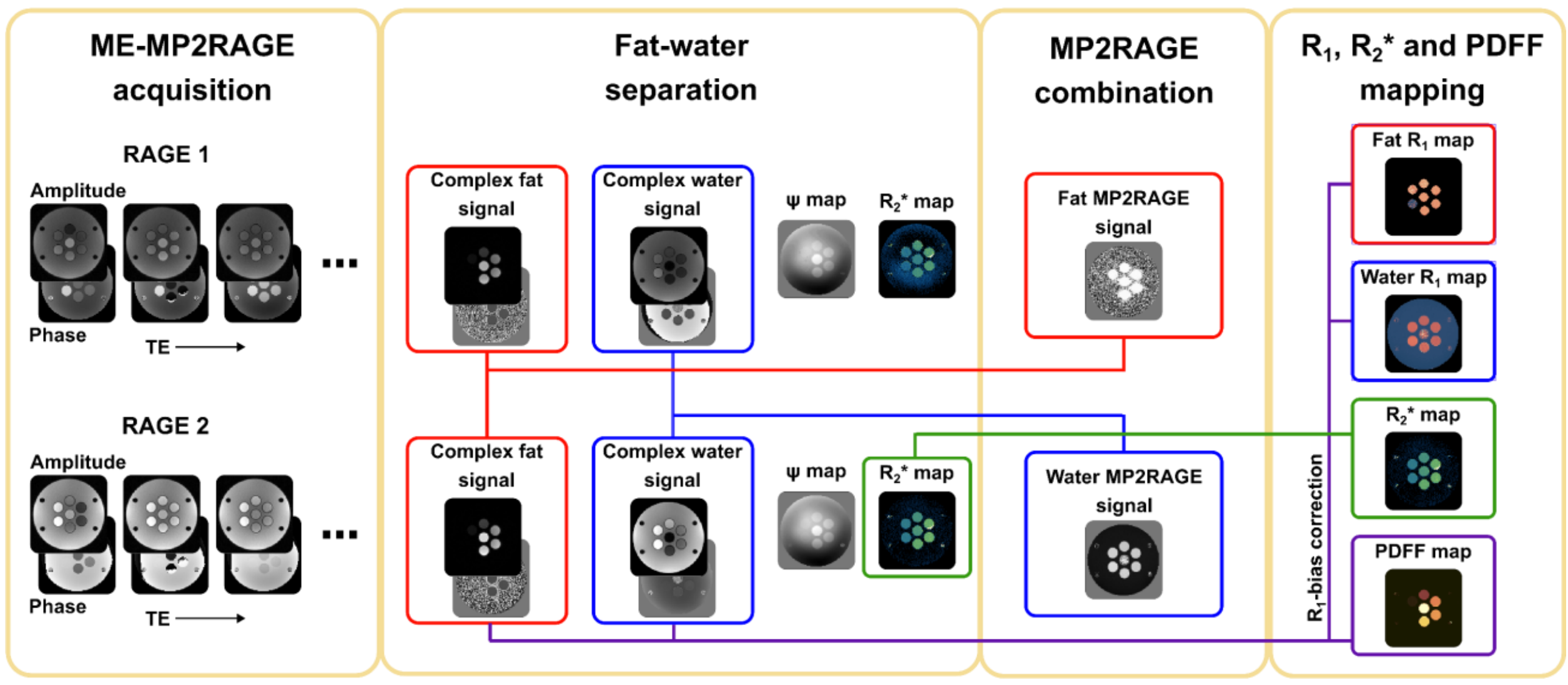

**Fig. 1:** Schematic representation of joint mapping of $R_{1f}$, $R_{1w}$, $R_2^*$, and PDFF using the ME-MP2RAGE sequence combined with CSE fat-water separation.

## 3 Methods

### 3.1 Sequence parameter optimization

Cramér-Rao lower bound (CRB) theory was used to optimize the ME-MP2RAGE sequence for mapping $R_{1f}$, $R_{1w}$, $R_2^*$, and SFF. Optimization was done independently for the magnetization prepared (MP-optimization) and multi-echo readout (ME-optimization) portions of the sequence. For each parameter, we calculated the value-to-noise-ratio (VNR) [54]—where V represents $R_{1f}$, $R_{1w}$, $R_2^*$, or SFF—across a range of sequence parameters and $R_{1f}$, $R_{1w}$, $R_2^*$, and SFF values. Subsequently, VNR was normalized by the signal-to-noise-ratio (SNR), and the dimensionless metric VNR/SNR was used to quantify the noise performance of the mapping technique [54]. For the calculations, the noise standard deviation of the signals in the RAGE blocks was set to yield a desired SNR, which was defined as the ratio of the magnitude signal in the first echo of the second RAGE block and the noise standard deviation. Finally, optimal sequence parameters were selected by maximizing the minimum VNR/SNR calculated across $R_{1f}$, $R_{1w}$, $R_2^*$, and SFF values.

The derivation of VNR/SNR is presented in Supplementary Information. For optimization, SFF was preferred over PDFF because the noise- and $R_1$-bias corrections made the calculation inaccurate, as explained in Supplementary Information. Validation of VNR/SNR calculations was done by comparing results with numerical simulations and phantom data. Further details about simulations and phantom experiments are presented in sections 3.2 and 3.3.

Our notation for sequence parameters follows prior convention [26]: $TI_1$ and $TI_2$ are the time between the inversion pulse and the mid-point of each inner phase encoding loop; TR is the repetition time between excitations within the RAGE blocks; $\alpha_1$ and $\alpha_2$ are the excitation flip angles for each RAGE block; $TR_{MP2RAGE}$ is the repetition time for the outer phase encoding loop; $B_1^+$ represents the transmit $B_1$ field inhomogeneity; eff represents the inversion efficiency of the inversion pulse, and n is the number of excitations within each RAGE block. The echo spacing, the first echo time, and the number of echoes were labelled as $\Delta TE$, $TE_1$, and K, respectively.

**3.1.1 Magnetization-prepared (MP) optimization**

We optimized the three parameters that had the greatest impact on the MP2RAGE signal: $TI_1$, $TI_2$, and $TR_{MP2RAGE}$. This was determined through a qualitative analysis of the theoretical MP2RAGE LUT [26] as function of $T_1$, $TI_1$, $TI_2$, $\alpha_1$, $\alpha_2$, $TR_{MP2RAGE}$, TR, n, and eff, as summarized in Fig C.1. We also analyzed the impact of $B_1^+$ inhomogeneity in the LUT by considering $\pm 10\%$ and $\pm 20\%$ deviations in the prescribed flip angles $\alpha_1$ and $\alpha_2$ (Fig. C.2). These values were selected from a report at 3 T [26]. The parameter space considered for the calculations is summarized in Table B.1.

**3.1.2 Multi-echo (ME) optimization**

We optimized $TE_1$ and $\Delta TE$, and studied the noise performance of protocols with different number of echoes and readout gradient schemes (unipolar and bipolar). Calculations were performed for four protocols identified as ME-MP2RAGE I, II, III, and IV. Protocols I and II have 3 and 6 echoes with unipolar readout gradients. Three echoes represent the minimum required for fat-water separation when estimating two complex signals corresponding to the fat and water compartments, together with two real-valued parameters, ($\psi$ and $R_2^*$). Protocols III and IV have 6 and 10 echoes with bipolar readout gradients. Six echoes represent the minimum required for the bipolar fat-water separation approach used in this work [55].

**3.2 Numerical simulations for evaluating accuracy in the presence of noise**

Using Monte Carlo (MC) simulations, we evaluated the theoretical accuracy of the protocols ME-MP2RAGE I, II, III, and IV. For the simulations, synthetic signals were generated using the model described in Eq. 1. Additive white gaussian noise was added to the real and imaginary parts of the signal using the same SNR definition as in VNR/SNR calculation. Simulations were repeated for 3000 noisy realizations. Fat-water separation for unipolar readouts was performed with a graph-cut based-technique [56]. For bipolar datasets, fat-water separation was performed with a modified version of the same technique used for unipolar readouts. This technique uses odd- and even-echo datasets to correct phase and amplitude errors induced by bipolar gradients [57]. The signed bias (=estimated parameter – ground truth), relative signed bias (=100 × [estimated parameter – ground truth] / ground truth), and their interquartile range (IQR) were used to quantify the accuracy of the estimates.

We performed simulations to evaluate the accuracy of ME-MP2RAGE IV for a wide range of $R_{1f}$, $R_{1w}$, $R_2^*$, and PDFF values for a moderate SNR=30. This protocol was selected as a representative example of the best accuracy achievable with the ME-MP2RAGE sequence used for multiparametric mapping. We also evaluated all protocols across different PDFF and SNR values. Finally, we used simulations to validate the calculation of VNR/SNR by comparing the VNR/SNR for $R_{1f}$, $R_{1w}$, $R_2^*$, SFF, and PDFF derived from CRB-based calculations and MC simulations. For validation, IDEAL with simultaneous $R_2^*$ estimation [58] and corrections for bipolar-induced effects [46], was used for fat-water separation since the least-squares estimation is the maximum likelihood estimator for additive white gaussian noise [59]. The parameters for all simulations are presented in Table B.1 and were selected according to ME- and MP-optimization.

**3.3 Phantom experiments**

Phantom experiments were used to evaluate the proposed technique. A custom phantom was assembled with 7 vials (50 mL centrifuge tubes, Corning) placed inside a large cylindrical enclosure (Magphan SMR170, The Phantom Laboratories). The vials inside the phantom were: 1 vial filled with 3% by weight agar gel prepared by mixing deionized distilled water, agar powder (MilliporeSigma Canada Ltd), and gadolinium-based contrast agent (GBCA; Gadovist 1.0 M, Bayer Healthcare); 5 vials filled with fat-water emulsions; and 1 vial filled with pure peanut oil (JVF Canada Inc). The fat-water emulsions were prepared following a published protocol [60] by mixing deionized distilled water, agar powder, peanut oil, GBCA, two non-ionic surfactants (Tween 20, MilliporeSigma Canada Ltd and Span 80, MilliporeSigma Canada Ltd), and sodium benzoate as a preservative (MilliporeSigma Canada Ltd). For the single vial filled with agar, the GBCA concentration was 0.20 mM. For the 5 vials containing fat-water emulsions, the nominal fat volume fractions were 5%, 25%, 50%, 60%, and 75% and GBCA concentrations were 0.19, 0.15, 0.10, 0.08, and 0.05 mM, respectively. The large phantom compartment was filled with deionized distilled water mixed with GBCA (0.03 mM) and table salt (85 mM). A diagram of the phantom is presented in Fig. C.3.

Data acquisition for all phantom experiments was performed using a 3 T scanner (Prisma, Siemens Healthineers) with the vendor-provided 20-channel head/neck coil. Phantom temperature was measured before (20.2 °C) and after (20.7 °C) the experiments by introducing a thermometer (long-stem thermometer, Traceable) inside the large phantom compartment. Sequence parameters were selected according to VNR/SNR calculations and are detailed in Table B.1. We collected data for ME-MP2RAGE I-IV but analysis focuses on protocols II and IV which performed the best in MC simulations. Fat-water separation for unipolar and bipolar datasets was performed with the same graph-cut based techniques used for numerical simulations [56,57]. Relaxation rate maps are presented using recently proposed colormaps [61], while other maps are presented using perceptually uniform colormaps [62].

The proposed technique was validated by comparison with reference measurements. An inversion recovery spin-echo (IR-SE) experiment was used to measure global $R_1$, which served as the reference for $R_{1f}$ in 100% oil and $R_{1w}$ in 100% agar and doped water. A multi-echo 3D fast low angle shot (3D FLASH) sequence was used as reference for $R_2^*$ and PDFF. Sequence parameters for IR-SE and 3D FLASH measurements are detailed in Table 1. Global $R_1$ maps were generated using publicly-available code [63], while $R_2^*$ and PDFF were obtained from fat-water separation with the same graph-cut based-technique [56] used for the ME-MP2RAGE data. Agreement between the reference and ME-MP2RAGE measurements was quantified using the intraclass correlation coefficient (ICC; "poor"<0.5, "moderate"=0.5–0.75, "good"=0.75–0.9, "excellent">0.9)[64]. For analysis, the maps were evaluated in 8 cylindrical regions of interest (ROI) each with a diameter of 4 voxels and spanning 5 contiguous slices. We placed 7 ROIs inside the vials and 1 ROI in the large phantom compartment as illustrated in Fig. C.3.

**Table 1:** Sequence parameters for ME-MP2RAGE experiments and reference experiments. All ME-MP2RAGE acquisitions had a GRAPPA acceleration factor of 2 in the outer loop, no partial Fourier, no flow compensation, FOV=208×208×188mm$^3$, and 4 min 28 s scan time. rBW = readout bandwidth, n = number of excitations in each RAGE block, K = total number of echoes.

**ME-MP2RAGE experiments**

| Protocol | $TR_{MP2RAGE}$ [s] | $TI_1$, $TI_2$ [s] | TR [ms] | $\alpha_1$, $\alpha_2$ [°] | n | $TE_1$ [ms] | ΔTE [ms] | rBW [Hz/px] | K | Voxel size [mm$^3$] |
|---|---|---|---|---|---|---|---|---|---|---|
| **I** | 4 | 0.6, 2.2 | 5.5 | 4, 4 | 104 | 1.01 | 1.63 | 1720 | 3, unipolar | 2.0×2.0×2.0 |
| **II** | 4 | 0.6, 2.2 | 10.4 | 4, 4 | 104 | 1.01 | 1.64 | 1720 | 6, unipolar | 2.0×2.0×2.0 |
| **III** | 4 | 0.6, 2.2 | 6.9 | 4, 4 | 104 | 1.01 | 0.93 | 1720 | 6, bipolar | 2.0×2.0×2.0 |
| **IV** | 4 | 0.6, 2.2 | 10.6 | 4, 4 | 104 | 1.01 | 0.93 | 1720 | 10, bipolar | 2.0×2.0×2.0 |

**Fat-water separation reference experiment**

| Protocol | TR [ms] | α [°] | $TE_1$ [ms] | ΔTE [ms] | rBW [Hz/px] | K | Voxel size [mm$^3$] |
|---|---|---|---|---|---|---|---|
| **3D FLASH** | 24 | 3 | 1.09 | 1.82 | 1700 | 6 (unipolar) | 1.7×2.0×1.7 |

**$R_1$ global mapping reference experiment**

| Protocol | TR [ms] | TIs [ms] | TE [ms] | rBW [Hz/px] | Voxel size [mm$^3$] |
|---|---|---|---|---|---|
| **IR-SE** | 5000 | 50, 100, 200, 400, 800, 1300, 2000, 4000 | 4.8 | 797 | 1.7×1.7×5.0 |

Differences in the variance of the estimates were tested for statistical significance with a two-way Levene's test to identify the most precise protocol. Analysis was done separately for $R_{1f}$, $R_{1w}$, $R_2^*$, and PDFF maps and the level of statistical significance was selected as α=0.05. The data was extracted from the ROIs shown in Fig. C.3. For the statistical test, the independent variables were the protocols (2 levels: ME-MP2RAGE II and IV) and the ROIs (8 levels). In case of the rejection of the null hypothesis when comparing protocols, the most precise protocol was determined based on the pooled standard deviation $\hat{\sigma} = \sqrt{\sum_{ROI=1}^{8} \frac{\sigma_{ROI}^2}{8}}$.

We compared VNR/SNR estimates from phantom experiments and CRB-based calculations. The experimental VNR was estimated as the ratio of the mean and standard deviations of $R_{1f}$, $R_{1w}$, $R_2^*$, and PDFF evaluated in the phantom ROIs. Meanwhile, SNR in each ROI was approximated by the mean and standard deviation of the magnitude signal in the RAGE2 block.

### 3.4 In vivo experiments

The proposed multiparametric mapping technique was evaluated in vivo in the lower leg and neck of a healthy volunteer (male, 24 years old). The volunteer gave informed consent and was imaged using the same 3 T MR scanner as for the phantom experiments. The lower leg dataset was acquired using the vendor's 15 channel transmit/receive knee coil. The neck dataset was acquired using the vendor's 20-channel head/neck coil. In vivo experiments used ME-MP2RAGE protocols II and IV, with minor adjustments for each anatomy. The lower leg dataset acquisition featured number of excitations n=72, FOV=160×144×188 $mm^3$, and scan time 3 min 36 seconds. The neck dataset included n=128, FOV=208×256×188 $mm^3$, and scan time 4 min 25 seconds. For the neck scans, the shortest achievable $TI_1$ values were 0.688 s (protocol II) and 0.702 s (protocol IV). All other sequence parameters followed those listed in Table 1. Quality of the fit for each protocol was evaluated with the normalized root-mean-square-error (NRMSE), the RMSE normalized by the magnitude signal in the first echo in the second RAGE block.

In vivo quantitative maps were analysed with ROIs placed in a variety of tissues. In the leg, the analysis included three manually created circular ROIs in yellow bone marrow (YBM), subcutaneous adipose tissue (SAT), and muscle. In the neck, the analysis included SAT and muscle ROIs. Due to the small size of SAT in the neck datasets, the ROI was created by thresholding voxels with PDFF>0.60 within a manually selected rectangular region. ROI placement is presented in Fig. C.4. Results for protocols ME-MP2RAGE II and IV were summarized by presenting the median and standard deviation. The voxel-wise measurements in each ROI for both protocols were compared using Bland-Altman (BA) plots.

# 4 Results

## 4.1 Sequence parameter optimization

### 4.1.1 Magnetization-prepared (MP) optimization

Optimal inversion times were selected as $TI_1$=0.6 s and $TI_2$=2.2 s. Fig. 2 presents VNR/SNR heat maps for $R_{1f}$, $R_{1w}$, $R_2^*$, and SFF as function of $TI_1$ and $TI_2$. This figure shows that inversion time selection has more impact on $R_{1f}$ and $R_{1w}$ compared to $R_2^*$ and SFF. Consequently, $TI_1$ and $TI_2$ selection focused on the noise performance of $R_{1f}$ and $R_{1w}$. Using short $TI_1$ and $TI_2$ increases $R_{1f}$NR/SNR. However, the minimum $TI_1$ and $TI_2$ are limited by physical constraints ($TI_1$ and $TI_2$ need to be long enough to fit each RAGE block within $TR_{MP2RAGE}$). The value of $TI_2$ also limits the smallest $R_{1w}$ that can be mapped since it shifts the inflection point of the MP2RAGE LUT. Fig. C.1 shows the MP2RAGE LUTs and their dependence on sequence parameters. The combination of $TI_1$=0.6 s and $TI_2$=2.2 s maximizes the minimum $R_{1w}$NR/SNR, enables mapping for $R_{1w}$<3.5 s (which covers the relaxation of most tissues at 3 T [65]), and offers near-optimal performance for $R_{1f}$.

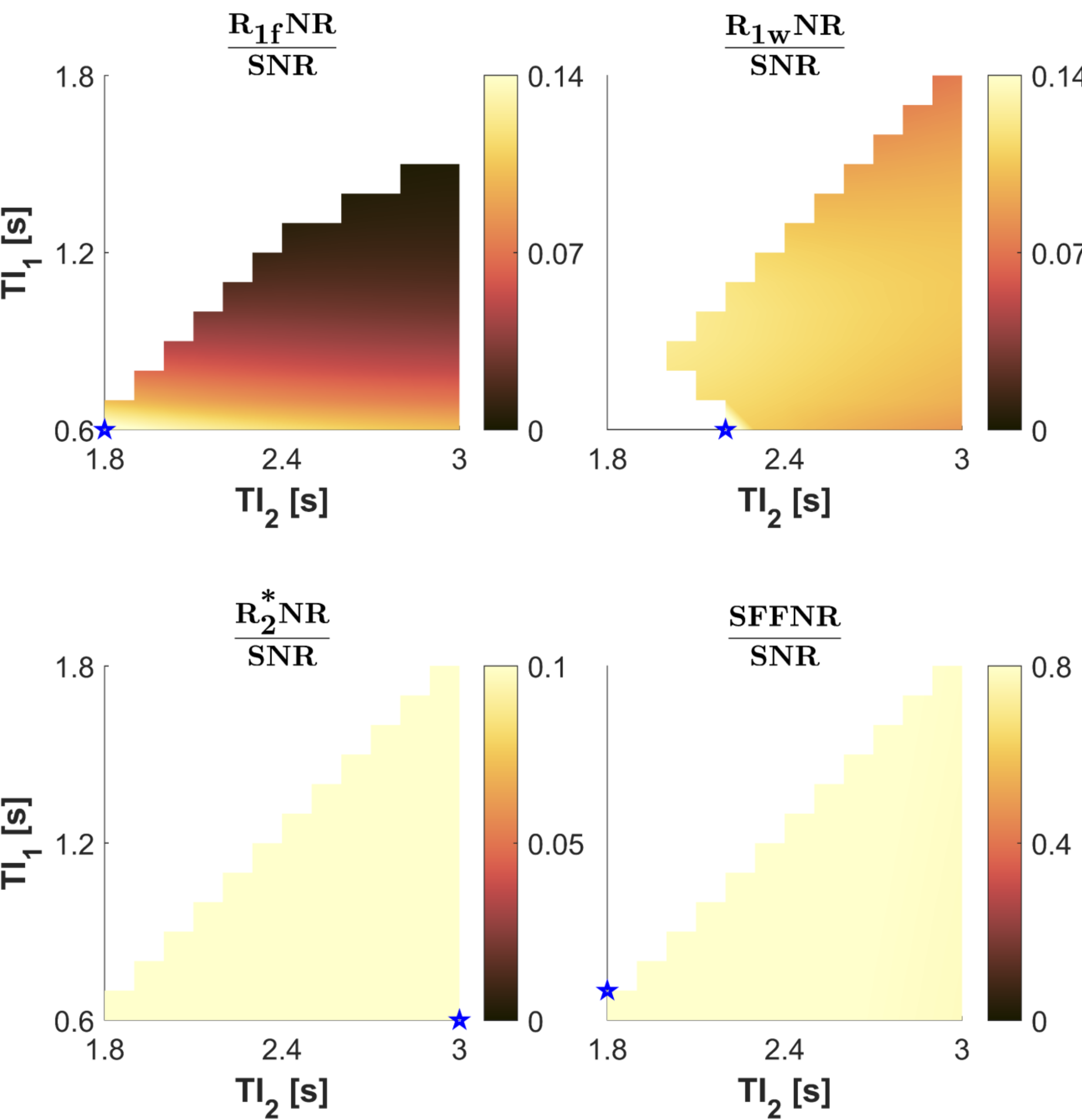

**Fig. 2:** Minimum VNR/SNR heat maps as function of $TI_1$ and $TI_2$ for $R_{1f}$, $R_{1w}$, $R_2^*$, and SFF. Blue markers: $TI_1$ and $TI_2$ combinations that maximize VNR/SNR in each map. $TI_1$ and $TI_2$ combinations that yield unphysical scenarios are omitted from the maps. $TI_1$=0.6 s was the shortest possible value for the considered n (number of excitations within each RAGE block). $TI_1$ and $TI_2$ selection has a high impact for $R_{1f}$ and $R_{1w}$, and a low impact for $R_2^*$ and SFF mapping.

Based on a qualitative analysis of the MP2RAGE LUTs (Fig. C.1), it was determined that: 1) the TI selection is the main factor to modify accuracy and precision of $R_{1f}$ and $R_{1w}$ (agreeing with VNR/SNR calculations), 2) that small flip angles are convenient for improved $R_{1f}$ mapping, and 3) $TR_{MP2RAGE}$ modifies the range of $R_1$ values for which the function is monotonic and linear, determining the $R_1$ range that can be measured with high accuracy and precision [26].

We selected the optimal $TR_{MP2RAGE}$ of 4 s. In general, VNR/SNR of $R_{1f}$ and $R_{1w}$ increased with increasing $TR_{MP2RAGE}$, while it decreased for SFF and $R_2^*$. This observation remained consistent for different combinations of $TI_1$, $TI_2$, $R_{1f}$, and $R_{1w}$ (results not shown). $TR_{MP2RAGE}$=4 s is relatively short, helping to reduce scan time while maintaining a good balance between noise performance for all parameters and the accuracy and precision of $R_1$.

Both flip angles, $\alpha_1$ and $\alpha_2$, were set to 4°. For our experiments, we identified that flip angle selection strongly influenced sensitivity to $B_1^+$ inhomogeneities (Fig. C.2). Lower flip angles would have reduced $B_1^+$ sensitivity but also decreased SNR and narrowed the measurable $R_1$ range. With the selected $\alpha_1$ and $\alpha_2$, Fig. C.2 shows that in theory $\pm 20\%$ deviations from the prescribed flip angles resulted in errors below 3.0% for $R_{1f}$ and below 12.9% for $R_{1w}$.

**4.1.2 Multi-echo (ME) optimization**

For numerical simulations, phantom, and in vivo experiments, we selected $TE_1$=1.0 ms and $\Delta$TE=1.6 ms (unipolar readout) or $\Delta$TE=0.9 ms (bipolar readout). These values correspond to the minimum $TE_1$ and $\Delta$TE achievable in experiments presented in this paper. Fig. 3 shows VNR/SNR heat maps for $R_{1f}$, $R_{1w}$, $R_2^*$, and SFF as function of $TE_1$ and $\Delta$TE. The blue markers in this figure indicate the $TE_1$ and $\Delta$TE pairs that maximize VNR/SNR. The minimum $TE_1$ and $\Delta$TE offer optimal or near optimal noise performance across all parameters for ME-MP2RAGE II and IV.

Increasing the number of echoes inside each RAGE block increased VNR/SNR for all estimates and bipolar readout gradients enabled more echoes within each RAGE block without changing $TI_1$, $TI_2$, $TR_{MP2RAGE}$. Fig. 3 shows that ME-MP2RAGE IV (10 echoes, bipolar) had the greatest VNR/SNR for all estimates. ME-MP2RAGE IV with $TE_1$=1.0 ms and $\Delta$TE=0.9 ms offered 38%, 29%, 26%, and 35% higher VNR/SNR for $R_{1f}$, $R_{1w}$, $R_2^*$, and SFF, compared to ME-MP2RAGE II with $\Delta$TE=1.6 ms. ME-MP2RAGE II and III (6 echoes unipolar and bipolar, respectively) had comparable VNR/SNR except near $\Delta$TE=1.1 ms where protocols with bipolar readouts presented a steep VNR/SNR decrease for $R_{1f}$, $R_{1w}$, and SFF. ME-MP2RAGE I (3 echoes unipolar) offered the lowest VNR/SNR.

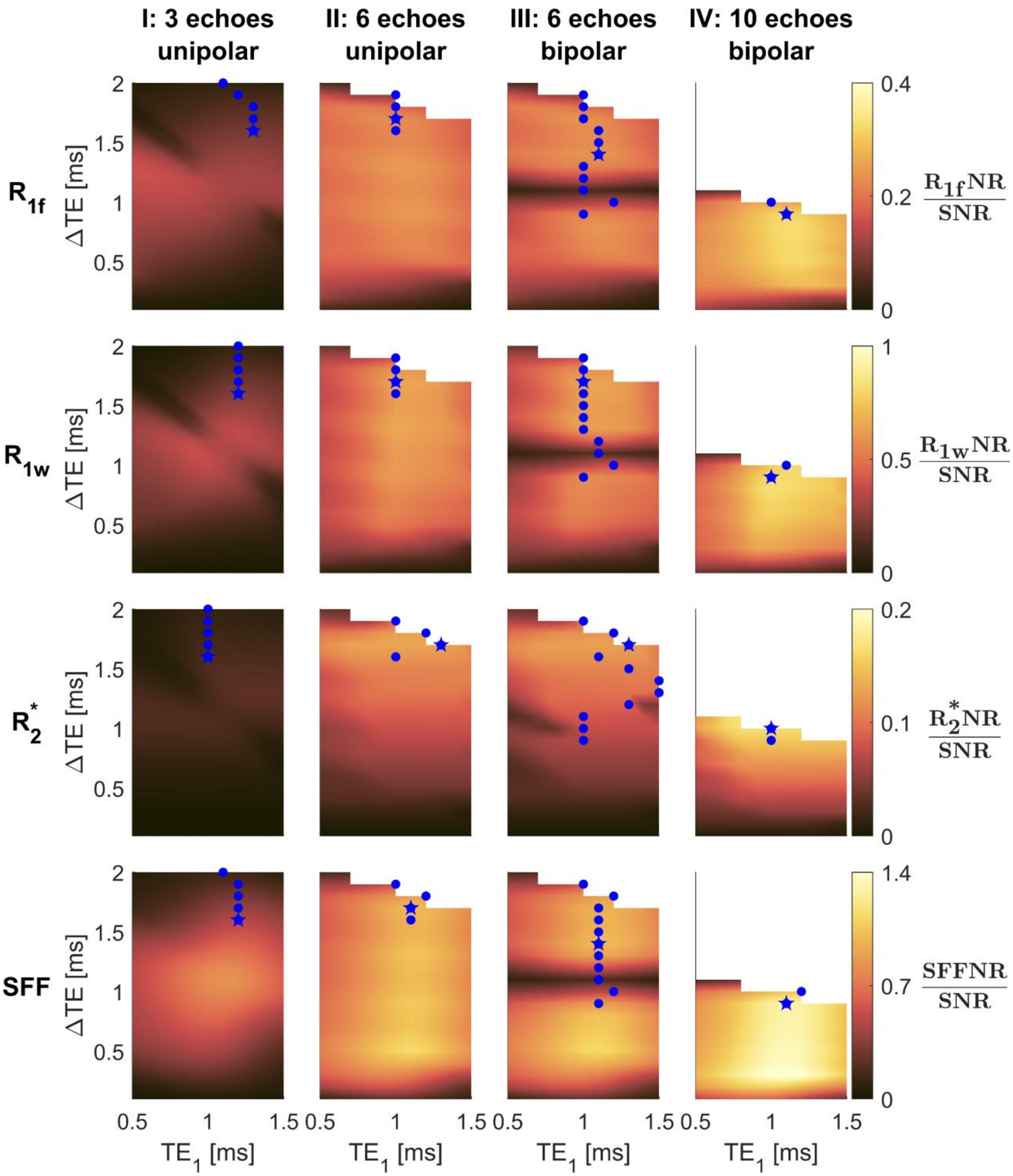

**Fig. 3:** Minimum VNR/SNR heat maps as function of $TE_1$ and ΔTE. Rows: Parameters $R_{1f}$, $R_{1w}$, $R_2^*$, and SFF. Columns: ME-MP2RAGE I-IV. Blue markers: The circles indicate $TE_1$ that maximizes VNR/SNR for each ΔTE, and the star indicates the pair with the maximum VNR/SNR among all marked pairs. $TE_1$ and ΔTE combination that yield unphysical scenarios are omitted from the maps. For all estimates, ME-MP2RAGE IV has the highest VNR/SNR.

### 4.2 Numerical simulations for evaluating accuracy in the presence of noise

In MC simulations with SNR = 30, ME-MP2RAGE IV accurately estimated $R_{1f}$, $R_{1w}$, $R_2^*$, and PDFF. Fig. 4 a), b), and c) show relative bias (mean and IQR) across a broad range of relaxation rates, with mean relative bias $< 3\%$ for $R_{1f} < 5.00$ s$^{-1}$ and $R_{1w} > 0.30$ s$^{-1}$, and $< 1\%$ for $R_2^* \in [10,100]$ s$^{-1}$. Fig. 4 d) highlights SFF inaccuracy due to $R_1$- and noise-bias: $R_1$-bias varied with $R_{1w}$ and peaked near PDFF$= 0.50$, while noise-bias affected values near PDFF 0 or 1. Fig. 4 e) shows that after $R_1$-bias correction (Supplementary Information), estimates remained inaccurate near PDFF 0 or 1. Finally, Fig. 4 f) shows that magnitude discrimination (as implemented in Supplementary Information) eliminated this effect and made PDFF accurate across PDFF $\in [0,1]$ with mean bias $< 0.003$. Comparable accuracy was observed for ME-MP2RAGE II (results not shown).

Increasing the number of echoes in each RAGE block boosted the accuracy of the estimates. Fig. C.5 shows that for unipolar and bipolar readout schemes, increasing the number of echoes from the minimum number required for multiparametric mapping substantially improved accuracy across PDFF and SNR values. ME-MP2RAGE II and IV offered comparable accuracy except at low SNR where ME-MP2RAGE IV slightly boosted accuracy of the estimates.

CRB-based calculations of VNR/SNR matched values derived from MC simulations for $R_{1f}$, $R_{1w}$, $R_2^*$, and SFF (Fig. C.6). Results did not match for PDFF, which was the rationale for optimizing sequence parameters based on SFF.

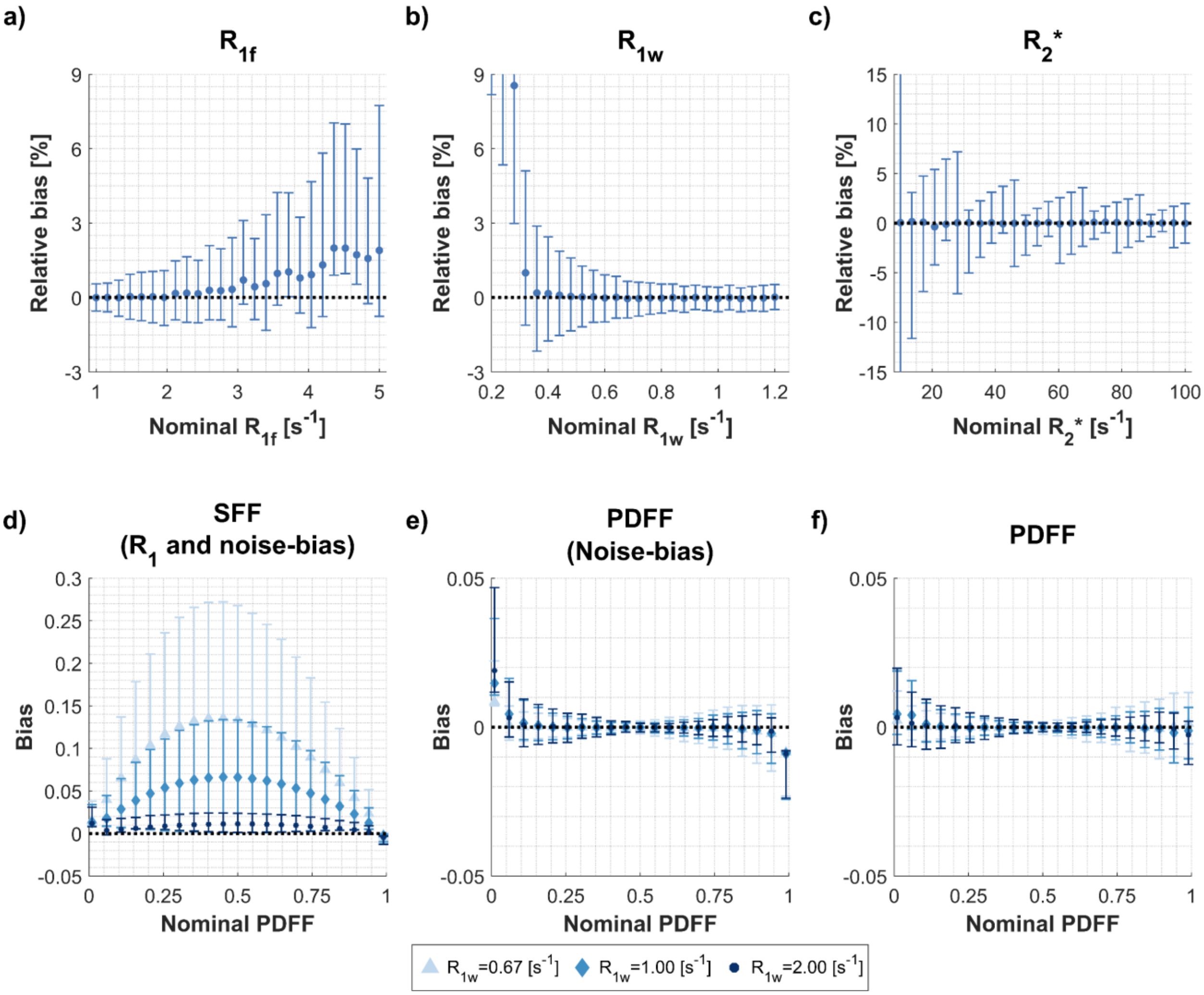

**Fig. 4:** MC simulation results for accuracy of the ME-MP2RAGE IV protocol (10 echoes, bipolar). Top row: Relative bias for a) $R_{1f}$, b) $R_{1w}$, and c) $R_2^*$ as a function of their corresponding nominal value. Bottom row: Bias for e) SFF, f) PDFF without noise-bias correction, and PDFF with $R_1$- and noise-bias corrections (calculations were done for different $R_{1w}$ values). Markers represent the mean, and error bars the IQR. The ME-MP2RAGE sequence can accurately estimate $R_{1f}$, $R_{1w}$, $R_2^*$, and PDFF.

### 4.3 Phantom experiments

ME-MP2RAGE measurements agreed with IR-SE and 3D FLASH measurements. Fig. 5 and 6 compare ME-MP2RAGE, IR-SE, and 3D FLASH maps with violin plots and linear plots, respectively. Analysis focused on protocols ME-MP2RAGE II and IV, as the other protocols produced estimates with a high noise level, particularly evident in the relaxation rate maps. Fig. C.7 shows maps for all protocols and highlights the poor noise performance of protocols ME-MP2RAGE I and III.

Excellent agreement with IR-SE was observed in the three ROIs used for comparison, where the global $R_1$ values from IR-SE could be meaningfully compared with ME-MP2RAGE-derived $R_{1w}$ or $R_{1f}$. Specifically, $R_{1w}$ estimates from ROIs 1 and 2 (100% doped distilled water and agar, respectively) and $R_{1f}$ estimates from ROI 8 (100% oil) were compared with the corresponding global $R_1$ values from IR-SE. Fig. 5 a) and b) show that ME-MP2RAGE II and IV produced comparable $R_{1f}$ and $R_{1w}$ estimates, both closely matching the global $R_1$ values obtained with IR-SE in these ROIs. This agreement is further supported by Fig. 6, which shows $ICC > 0.9$ for both protocols. In fat-water emulsion vials, $R_{1f}$ remained relatively constant across nominal fat volume fractions, whereas $R_{1w}$ decreased slightly with increasing fat volume fraction (Fig. 5).

For $R_2^*$ and PDFF, violin plots and linear plots also show an excellent agreement ($ICC>0.9$) between ME-MP2RAGE II and IV and FLASH 3D measurements (Fig. 5 and 6). $R_2^*$ maps presented the largest difference between protocols II and IV compared to other parameters. These discrepancies can be explained by the differences in readout and $\Delta TE$. On the other hand, PDFF measurements remained accurate due to corrections for $R_1$- and noise-bias, which were the most impactful in vials of nominal fat fraction 0.50 and 1.00 (Fig. C.8).

ME-MP2RAGE IV was the most precise protocol. The pooled variance $\hat{\sigma}$ was lower for ME-MP2RAGE IV compared to II for all quantitative maps. Levene's test showed a statistically significant effect for the factor protocol for $R_{1f}$ ($p=0.03$), $R_2^*$ ($p<0.001$), and PDFF ($p<0.001$). There were no significant effects for $R_{1w}$ ($p=0.06$). For all quantitative maps, the effect of the factor ROI was significant ($p<0.001$ for all cases). The interaction effect was not significant, suggesting that precision differences between protocols do not depend on the nominal fat fraction.

Phantom experiments were also used to validate CRB-based calculation. Fig. C.9 shows that in phantoms, the theoretical VNR/SNR follows the trends measured in phantoms.

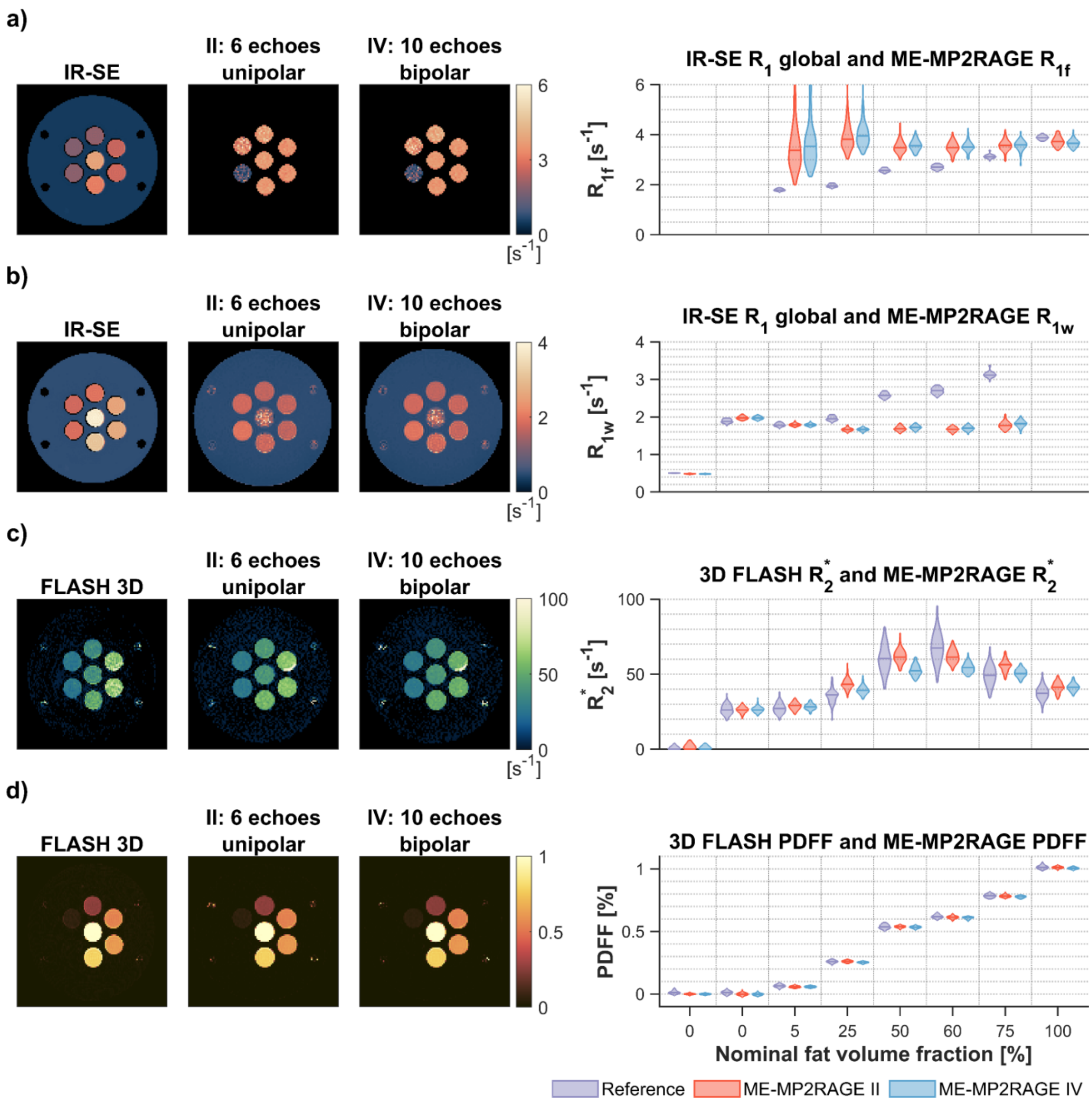

**Fig. 5:** Quantitative maps and violin plots for comparison of ME-MP2RAGE (protocols II and IV), IR-SE, and 3D FLASH measurements. Left: Quantitative maps of (a) $R_{1f}$, (b) $R_{1w}$, (c) $R_2^*$, and (d) PDFF. Right: Violin plots for $R_{1f}$, $R_{1w}$, $R_2^*$, and PDFF maps evaluated in ROIs 1-8.

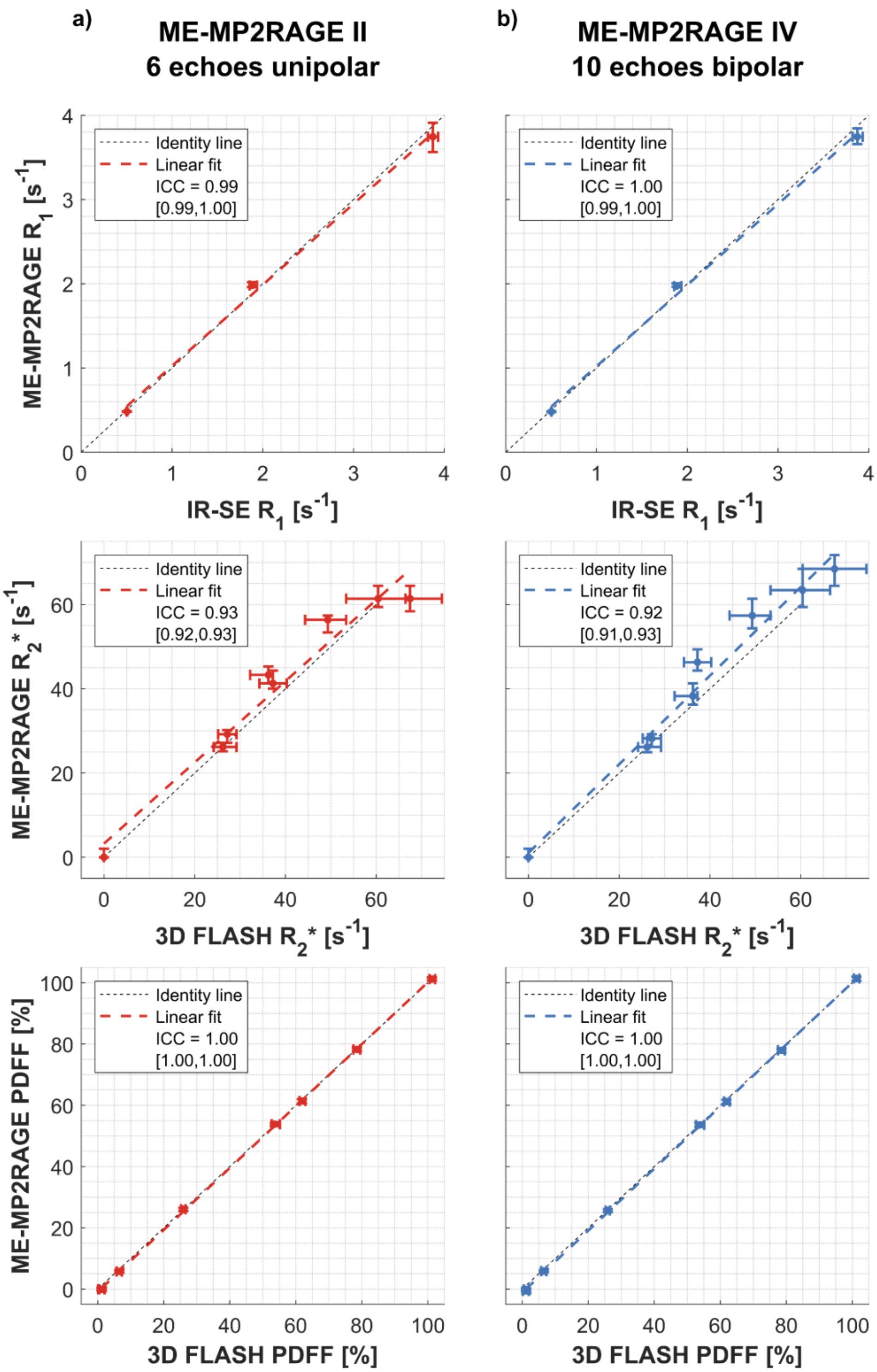

**Fig. 6:** Comparison of ME-MP2RAGE with IR-SE and 3D FLASH measurements in phantom experiments. Columns a) and b) present comparisons for protocols II and IV, respectively. ME-MP2RAGE $R_{1w}$ and $R_{1f}$ are compared with IR-SE $R_1$ global only in ROIs 1, 2, and 8, corresponding to 100% doped water, agar, and oil, where PDFF=0 or 1. ME-MP2RAGE and 3D-FLASH $R_2^*$ and PDFF values are compared across all vials. Markers and error bars represent the median and IQR, respectively.

### 4.4 In vivo experiments

Quantitative maps for the lower leg and neck yielded measurements consistent with literature values and variability attributable to tissue properties and anatomy. Fig. 7 shows $R_{1f}$, $R_{1w}$, $R_2^*$, PDFF, and NRMSE maps generated using protocols ME-MP2RAGE II and IV. Table 2 presents summary statistics for the maps evaluated in ROIs covering different tissues (shown in Fig. C.4). $R_{1f}$ measured in YBM and SAT (for lower leg and neck) was consistent across protocols II and IV, with median values within 4% of each other. These measurements are also consistent with values reported in literature [65]. Variations in SAT measurements between anatomies are in agreement with the observation that $R_{1f}$ varies slightly across adipose tissue [66]. $R_{1w}$ in muscle was also consistent across protocols and anatomies with a 4% difference in the median between protocols and anatomies, and in agreement with values reported in literature [65]. $R_2^*$ exhibited the largest IQRs and greatest variation across tissues and anatomies, consistent with the simulation results presented in this study, which identified $R_2^*$ as the most variable estimate. Meanwhile, variability between the neck and leg datasets can be explained due to $R_2^*$ dependence on external magnetic field inhomogeneities, which change with anatomy [67]. PDFF maps presented large variations between YBM and SAT, as well as across SAT sites, likely reflecting underlying differences between bone marrow and SAT [68], and SAT in the lower and upper-body [69]. NRMSE maps showed poor quality fits in cortical bone in the lower leg, and near the trachea and in vertebrae of the neck, leading to unreliable measurements in these regions.

The largest mean differences (±95% CI) between protocols ME-MP2RAGE II and IV were $-0.11\pm0.03$ $s^{-1}$ in lower-leg SAT for $R_{1f}$, $0.01\pm7\times10^{-3}$ $s^{-1}$ in neck muscle for $R_{1w}$, $9.3\pm1.2$ $s^{-1}$ in neck muscle for $R_2^*$, and $0.037\pm0.010$ in neck SAT for PDFF. Fig. 8 and Fig. C.10 show BA plots comparing measurements between the two protocols across ROIs for the two anatomies considered in the in vivo experiments. Overall, measurements in the neck showed larger mean differences between the protocols and wider limits of agreement (LoA). Comparisons of $R_{1w}$ in YBM and SAT, and $R_{1f}$ in muscle, were omitted due to high noise in measurements in tissue with extremely high or low PDFF, respectively.

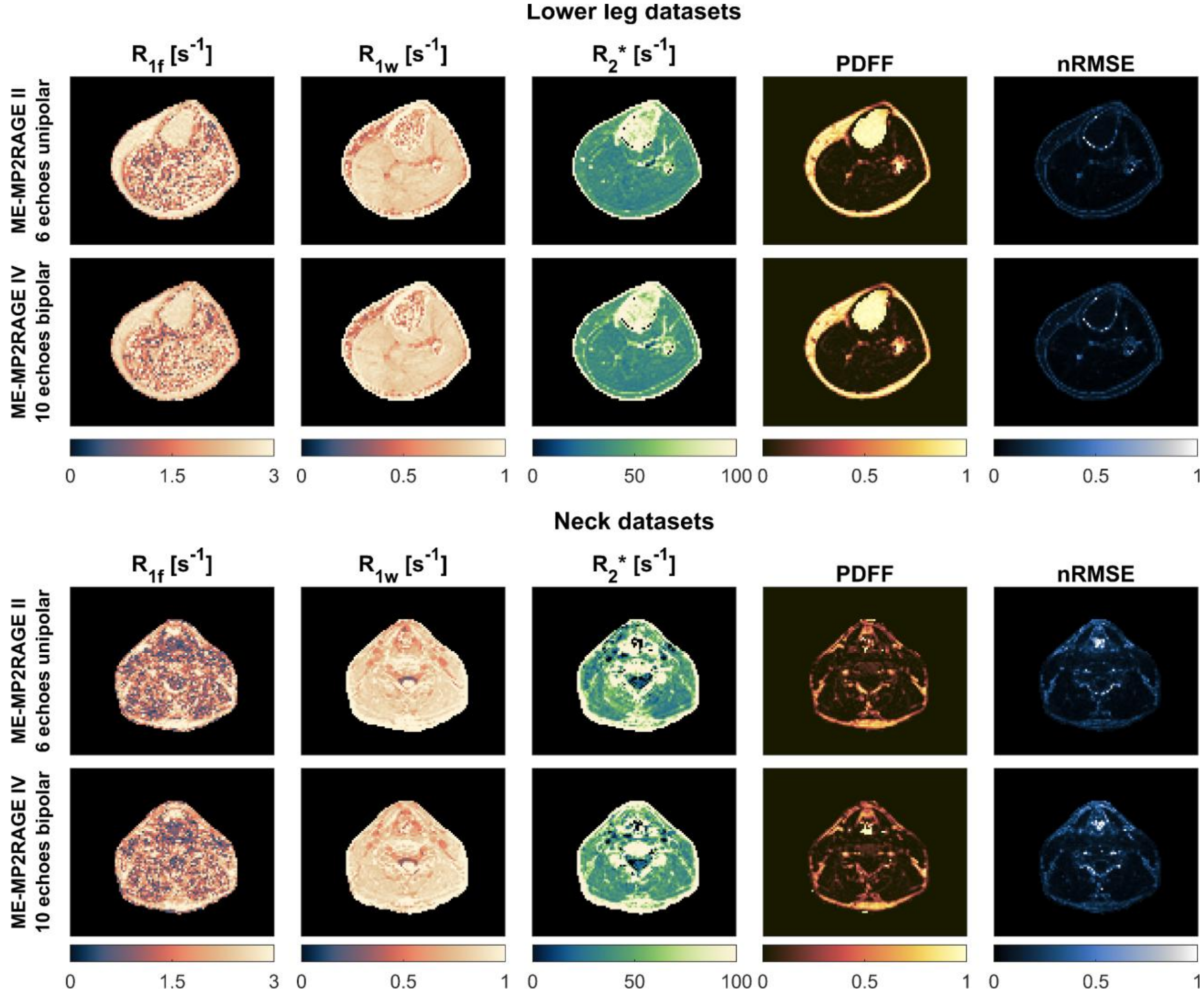

**Fig. 7:** In vivo pilot study using ME-MP2RAGE II and ME-MP2RAGE V protocols for the lower leg and neck. $R_{1f}$, $R_{1w}$, $R_2^*$, PDFF, and NRMSE maps are presented for each case.

**Table 2:** Median and IQR for $R_{1f}$, $R_{1w}$, $R_2^*$, and PDFF evaluated in different anatomical ROIs.

**Lower leg**

| | | $R_{1f}$ [$s^{-1}$] | $R_{1w}$ [$s^{-1}$] | $R_2^*$ [$s^{-1}$] | PDFF |
|---|---|---|---|---|---|
| Yellow bone marrow (YBM) | II: | 2.74 (0.17) | 0.79 (0.45) | 89.00 (15.12) | 0.97 (0.03) |
| | IV: | 2.66 (0.12) | 0.79 (0.37) | 90.00 (16.12) | 0.98 (0.03) |
| Subcutaneous adipose tissue (SAT) | II: | 2.79 (0.18) | 0.57 (0.21) | 42.50 (8.38) | 0.86 (0.11) |
| | IV: | 2.69 (0.15) | 0.59 (0.17) | 43.00 (8.38) | 0.88 (0.11) |
| Muscle | II: | 1.54 (0.91) | 0.88 (0.03) | 36.00 (4.50) | 0.02 (0.01) |
| | IV: | 1.87 (0.91) | 0.89 (0.03) | 36.00 (4.62) | 0.01 (0.01) |

**Neck**

| | | $R_{1f}$ [$s^{-1}$] | $R_{1w}$ [$s^{-1}$] | $R_2^*$ [$s^{-1}$] | PDFF |
|---|---|---|---|---|---|
| Subcutaneous adipose tissue (SAT) | II: | 2.98 (0.59) | 1.08 (0.25) | 84.50 (39.12) | 0.70 (0.11) |
| | IV: | 2.86 (0.59) | 0.95 (0.23) | 80.50 (28.00) | 0.73 (0.10) |
| Muscle | II: | 1.22 (0.94) | 0.86 (0.05) | 31.50 (7.62) | 0.02 (0.03) |
| | IV: | 1.27 (0.97) | 0.87 (0.04) | 42.50 (8.12) | 0.03 (0.02) |

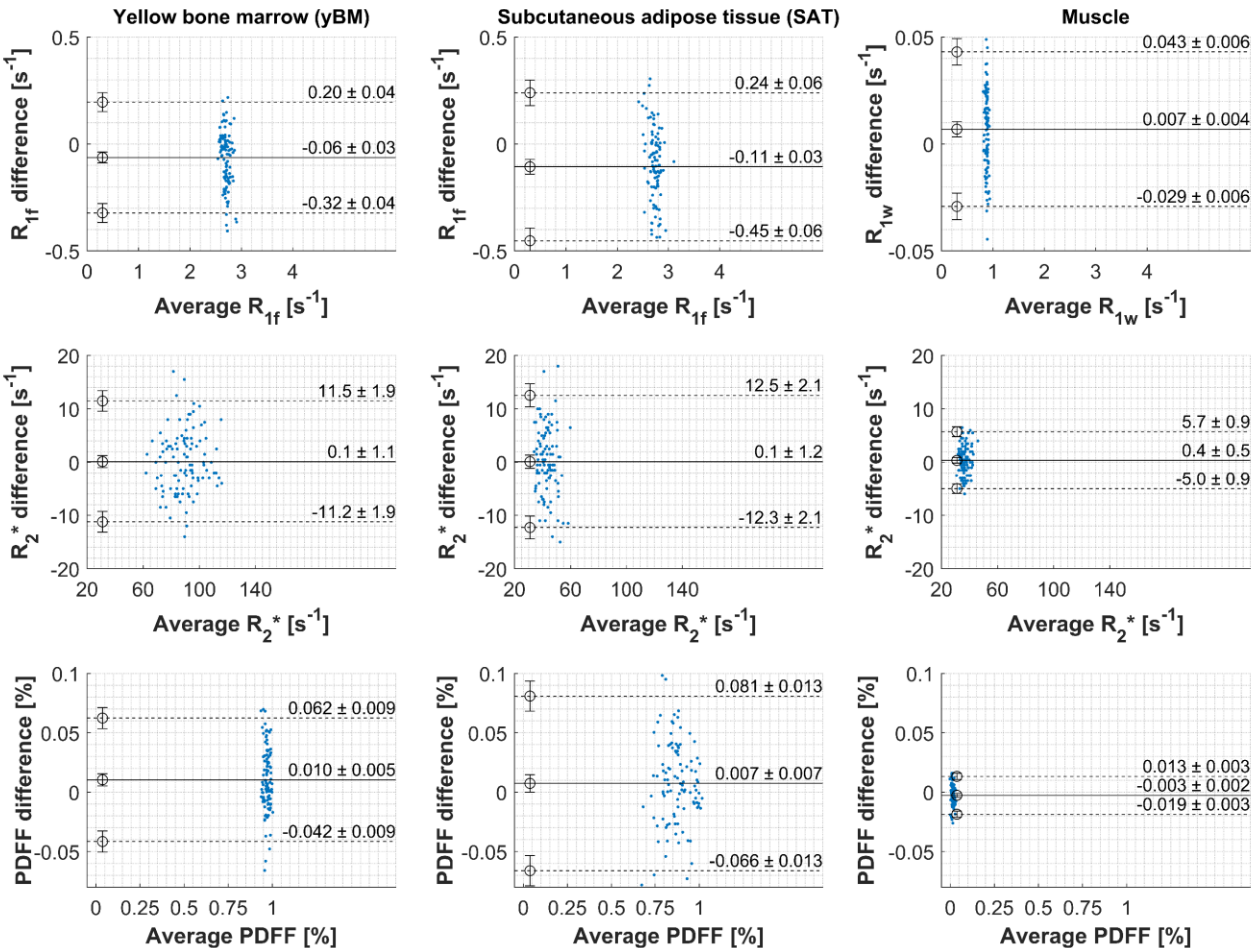

**Fig. 8:** Bland-Altman plots comparing ME-MP2RAGE II and IV for ROIs placed in the lower leg covering YBM, SAT, and muscle. Comparisons for $R_{1w}$ for YBM and SAT, and $R_{1f}$ for muscle are omitted, no significant differences were detected between protocols for these comparisons. Values above the solid line represent mean difference with 95% confidence intervals (CIs). Values above the dashed lines represent limits of agreement with 95% CIs.

## 5 Discussion

A technique to jointly map $R_{1f}$, $R_{1w}$, $R_2^*$, and PDFF using the ME-MP2RAGE sequence was presented. Sequence parameters were optimized through CRB-based calculations. Numerical simulations, and phantom and in vivo experiments were performed to test and validate this technique.

Literature has shown that MP2RAGE and IR-SE measurements agree with each other, making it an accurate alternative for global $R_1$ mapping [70]. Numerical simulations and phantom results presented in this paper show that the MP2RAGE-based relaxation mapping can be used to map fat- and water-specific $R_1$. Through numerical simulations, we demonstrated the accuracy of the proposed technique for a wide range of $R_{1f}$, $R_{1w}$, PDFF, and SNR values. $R_{1f}$ and $R_{1w}$ sensitivity to $B_1^+$ inhomogeneities could be further reduced through selection of pulse sequence parameters. In this work, we identified the RAGE flip angles as a critical parameter to mitigate $B_1^+$ induced bias. The flip angles used here (4°) provides acceptable $B_1^+$ bias while prioritizing SNR; our calculations suggest that slightly smaller flip angles (3°) may be preferred when $B_1^+$ robustness is the primary concern. In experiments in phantoms for vials with nominal PDFF=0 and 1, there was excellent agreement when comparing $R_{1f}$ and $R_{1w}$ with global $R_1$ from an IR-SE experiment. We did not compare our technique with a reference method for fat-water emulsions, as no widely accepted "gold-standard" exists for $R_{1f}$ and $R_{1w}$ mapping.

The proposed approach performs $R_2^*$ mapping as part of the fat-water separation step, unlike other approaches that use mono-exponential fitting of the magnitude data from the second RAGE block. The selected method for $R_2^*$ mapping eliminates the bias due to the presence of fat [20]. In numerical simulations and phantom experiments, $R_2^*$ was the least precise and least accurate parameter. In this work, suboptimal $R_2^*$ mapping was likely caused by the longest echo time not matching the $T_2$* value to be measured [47]. Further improvements in $R_2^*$ accuracy and precision may therefore be limited by the timing of the last echo, which is constrained by $TI_1$, $TI_2$, and $TR_{MP2RAGE}$.

Accurate PDFF mapping with the ME-MP2RAGE sequence was possible by removing $R_1$- and noise-bias. Given the sensitivity of $R_1$-bias to sequence parameters and tissue-specific $R_1$ values, the proposed correction is crucial for ensuring the reliable use of ME-MP2RAGE for PDFF mapping. At the same time, the noise-bias correction improves accuracy of the estimates at extremely high and low PDFF values.

Given the large number of sequence parameters to be optimized in the ME-MP2RAGE sequence and the goal of mapping multiple parameters, it is challenging to create an automatic optimization technique to maximize accuracy and precision of all maps [27]. We addressed this challenge by dividing the optimization problem into two independent parts, MP and ME-optimizations, based on their primary influence on $R_1$ mapping and fat-water separation, respectively. This approach works because MP2RAGE-based $R_1$ mapping is independent of differences in echo times and $R_2^*$ effects [26]. One caveat with this approach is that the performance predicted by CRB-based calculations applies to unbiased estimators. However, the fat-water separation method in the proposed approach uses a penalized maximum likelihood framework, meaning the estimators are not guaranteed to be unbiased. Moreover, unmodeled biophysical effects and artifacts introduce errors, preventing the performance predicted by CRB calculations from being achieved. Despite these limitations, CRB calculations were still deemed appropriate and effective in quantifying SNR-efficiency and to improve not only the precision but also the accuracy of the estimates [71,72]. Thus, CRB theory was used to optimize the sequence parameters and a ratio between VNR and SNR was used as a dimensionless metric for $R_{1f}$, $R_{1w}$, $R_2^*$, and PDFF mapping optimization, similar to how the number of signal averages (NSA) metric is used in fat-water separation optimization [73].

In simulations and phantom experiments, bipolar readout gradients allowed for a shorter ΔTE and more echoes per excitation, thus improving precision and accuracy, particularly in low SNR conditions, without compromising sequence parameter selection. Although not explored in this work, bipolar readout gradients can provide other advantages such as the flexibility of reducing voxel size by relaxing constraints on ΔTE and on the number of echoes. However, the use of bipolar readout gradients entails certain considerations. First, the potential improvements in accuracy and precision due to the increase in number of echoes are not as prominent due to the increase in degrees of freedom of the fat-water separation model because of the need for correction of effects induced by bipolar readout gradients [46,51,52]. Second, CRB calculations showed that there are limitations regarding ΔTE selection since VNR/SNR steeply decreased near ΔTE=1.1 ms, which matches with limitations reported in fat-water separation literature [46]. Third, to minimize the ΔTE and to eliminate field inhomogeneity and chemical shift-induced misregistration [51] caused by the bipolar readouts, we used a high readout bandwidth. Increasing the readout bandwidth has a detrimental effect in the SNR of the acquisition. This reduction in SNR is expected not to affect the accuracy and precision of $R_{1f}$, $R_{1w}$, $R_2^*$, and PDFF since the increased SNR efficiency brought by bipolar gradients compensates for the reduction in SNR[28].

The relatively slow nature of multi-shot acquisitions like the MP2RAGE sequence makes it sensitive to motion artifacts [74]. To showcase our technique, we selected anatomies in which there is low possibility of motion induced artifacts degrading the quality of the images and quantitative maps. It remains as future work to properly analyze this limitation as well as to establish options to manage motion induced artifacts. Some potential alternatives to achieve this is the use of navigator-based motion correction [74–77] or the use of non-Cartesian encoding schemes that are less sensitive to motion artifacts [78,79].

## 6 Conclusion

In this work, we presented an accurate and precise method for simultaneous quantification of $R_{1f}$, $R_{1w}$, $R_2^*$, and PDFF from a single acquisition with the ME-MP2RAGE sequence. We provided a comprehensive overview of this approach by developing a formalism for sequence parameter optimization and subsequently evaluated the expected accuracy and precision of the method through numerical simulations. Tests of the technique with different phantom and in vivo experiments showed that this technique is appropriate for deployment in future studies.

## CRediT authorship contribution statement

**Jorge Campos Pazmiño:** Conceptualization, Methodology, Software, Validation, Formal analysis, Investigation, Data curation, Writing – original draft, Writing review & editing, Visualization. Marc-Antoine Fortin: Conceptualization, Software, Resources, Writing review & editing. Véronique Fortier: Resources, Writing review & editing. Evan McNabb: Resources, Writing review & editing. Renée-Claude Bider: Methodology, Visualization. André van der Kouwe: Resources, , Writing review & editing, Supervision. Ives R. Levesque: Conceptualization, Methodology, Validation, Resources, Writing – review & editing, Supervision, Project administration, Funding acquisition.

## Data availability statement

Source code to map $R_{1f}$, $R_{1w}$, $R_2^*$, and PDFF with the ME-MP2RAGE sequence is shared on GitLab (https://gitlab.com/MPUmri/me_mp2rage_multiparametric_mapping.git). This code requires a working version of the ISMRM Fat-Water Separation Toolbox. Data from phantom experiments is shared on the Open Science Foundation (https://osf.io/wrmjg/).

## Acknowledgments

The authors acknowledge the developers of the ISMRM fat-water toolbox (http://www.ismrm.org/workshops/FatWater12/data.htm), Norma Ybarra for her contributions during phantom fabrication, and the Montreal General Hospital MRI Research Platform for support of experiments. Jorge Campos Pazmiño acknowledges funding provided from the Fonds de recherche du Québec – Nature et technologies (FRQNT; B2X - Bourse de doctorat en recherche). This project was funded by a Discovery Grant from the Natural Science and Engineering Research Council of Canada, *Chercheur boursier* and *Établissement de jeune chercheur* awards from the *Fond de recherche du Québec–Santé*, and by the Montreal General Hospital Foundation.

## A Supplementary Information

### A.1 Calculation of corrections for elimination of R1- and noise-bias in proton density fat fraction estimation

Eq. 1 in section 2 Theory of the main manuscript shows the two-component signal model used to represent the signal in each RAGE block. In that equation, $\mathrm{W_r}$ and $\mathrm{F_r}$ are the complex water and fat signal components independent of echo time in each RAGE block, with the subscript $\mathrm{r} = 1, 2$ representing the number of the RAGE block. Labelling the magnitude of the steady-state water and fat signal components for TE=0 as $\mathrm{w_r} = |\mathrm{W_r}|$ and $\mathrm{f_r} = |\mathrm{F_r}|$, for the RAGE1 and RAGE2 blocks, these terms are:

$$\mathrm{x_1} = \mathrm{B_1^-} \sin(\mathrm{B_1^+}\alpha_1)\,\mathrm{M_x}\left\{\left(-\mathrm{eff}\frac{\mathrm{m_{zx}^-}}{\mathrm{M_x}}\mathrm{E_{Ax}} + (1-\mathrm{E_{Ax}})\right)(\mathrm{E_{1x}}\cos(\mathrm{B_1^+}\alpha_1))^{\frac{n}{2}-1} + (1-\mathrm{E_{1x}})\left(\frac{1-(\mathrm{E_{1x}}\cos(\mathrm{B_1^+}\alpha_1))^{\frac{n}{2}-1}}{1-\mathrm{E_{1x}}\cos(\mathrm{B_1^+}\alpha_1)}\right)\right\} \quad \text{A.1}$$

$$\mathrm{x_2} = \mathrm{B_1^-} \sin(\mathrm{B_1^+}\alpha_2)\,\mathrm{M_x}\Bigg\{\Bigg\{\Bigg[\left(-\mathrm{eff}\frac{\mathrm{m_{zx}^-}}{\mathrm{M_x}}\mathrm{E_{Ax}} + (1-\mathrm{E_{Ax}})\right)(\mathrm{E_{1x}}\cos(\mathrm{B_1^+}\alpha_1))^{n} + (1-\mathrm{E_{1x}})\left(\frac{1-(\mathrm{E_{1x}}\cos(\mathrm{B_1^+}\alpha_1))^{n}}{1-\mathrm{E_{1x}}\cos(\mathrm{B_1^+}\alpha_1)}\right)\Bigg]\mathrm{E_{Bx}} + (1-\mathrm{E_{Bx}})\Bigg\}(\mathrm{E_{1x}}\cos(\mathrm{B_1^+}\alpha_2))^{\frac{n}{2}-1} + (1-\mathrm{E_{1x}})\left(\frac{1-(\mathrm{E_{1x}}\cos(\mathrm{B_1^+}\alpha_2))^{\frac{n}{2}-1}}{1-\mathrm{E_{1x}}\cos(\mathrm{B_1^+}\alpha_2)}\right)\Bigg\} \quad \text{A.2}$$

where $\mathrm{B_1^-}$ and $\mathrm{B_1^+}$ represent receive and transmit filed inhomogeneities, respectively; $\alpha_1$ and $\alpha_2$ are the flip angles for RAGE1 and RAGE2, respectively; eff is the inversion pulse efficiency, n is the number of excitations in each RAGE block. 'x' can be replaced by 'w' or 'f' for terms related to the water and fat signals, respectively. $\mathrm{M_x}$ is the density of protons from water or fat ($\mathrm{M_w}$ and $\mathrm{M_f}$, respectively), $\mathrm{m_{zx}^-}$ is the longitudinal magnetization under steady-state conditions, and $\mathrm{E_{Ax}} = e^{-\mathrm{TAR_{1x}}}$, $\mathrm{E_{Bx}} = e^{-\mathrm{TBR_{1x}}}$, $\mathrm{E_{Cx}} = e^{-\mathrm{TCR_{1x}}}$, and $\mathrm{E_{1x}} = e^{-\mathrm{TRR_{1x}}}$ for times TA, TB, TC, and TR as defined by Marques et al. [1] $\mathrm{m_{zx}^-}$ is expressed as:

$$\mathrm{m_{zx}^-} = \frac{\mathrm{M_x}\Bigg\{\Bigg\{\Bigg[\Bigg((1-\mathrm{E_{Ax}})(\mathrm{E_{1x}}\cos(\mathrm{B_1^+}\alpha_1))^{n} + (1-\mathrm{E_{1x}})\left(\frac{1-(\mathrm{E_{1x}}\cos(\mathrm{B_1^+}\alpha_1))^{n}}{1-\mathrm{E_{1x}}\cos(\mathrm{B_1^+}\alpha_1)}\right)\Bigg)\mathrm{E_{Bx}} + (1-\mathrm{E_{Bx}})\Bigg](\mathrm{E_{1x}}\cos(\mathrm{B_1^+}\alpha_2))^{n} + (1-\mathrm{E_{1x}})\left(\frac{1-(\mathrm{E_{1x}}\cos(\mathrm{B_1^+}\alpha_2))^{n}}{1-\mathrm{E_{1x}}\cos(\mathrm{B_1^+}\alpha_2)}\right)\Bigg\}\mathrm{E_{Cx}} + (1-\mathrm{E_{Cx}})\Bigg\}}{1+\mathrm{eff}(\cos(\mathrm{B_1^+}\alpha_1))^{n}(\cos(\mathrm{B_1^+}\alpha_2))^{n} e^{\mathrm{TR_{MP2RAGE}R_{1x}}}} \quad \text{A.3}$$

From Eq. A.2, it is possible to define the term $\mathrm{C_x}$:

$$\mathrm{C_x} = \mathrm{B_1^-} \sin(\mathrm{B_1^+}\alpha_2)\Bigg\{\Bigg\{\Bigg[\left(-\mathrm{eff}\frac{\mathrm{M_{zx}^-}}{\mathrm{M_x}}\mathrm{E_{Ax}} + (1-\mathrm{E_{Ax}})\right)(\mathrm{E_{1x}}\cos(\mathrm{B_1^+}\alpha_1))^{n} + (1-\mathrm{E_{1x}})\left(\frac{1-(\mathrm{E_{1x}}\cos(\mathrm{B_1^+}\alpha_1))^{n}}{1-\mathrm{E_{1x}}\cos(\mathrm{B_1^+}\alpha_1)}\right)\Bigg]\mathrm{E_{Bx}} + (1-\mathrm{E_{Bx}})\Bigg\}(\mathrm{E_{1X}}\cos(\mathrm{B_1^+}\alpha_2))^{\frac{n}{2}-1} + (1-\mathrm{E_{1x}})\left(\frac{1-(\mathrm{E_{1x}}\cos(\mathrm{B_1^+}\alpha_2))^{\frac{n}{2}-1}}{1-\mathrm{E_{1x}}\cos(\mathrm{B_1^+}\alpha_2)}\right)\Bigg\} \quad \text{A.4}$$

Given Eq. A.**Error! Reference source not found.** and the fact that $\frac{m_{zx}^{-}}{M_x}$ only depends on $R_{1x}$, it can be concluded that $C_x$ depends only on the terms $B_1^-$ and $R_{1x}$. Thus, $C_x$ can be used as a correction term to calculate a PDFF estimate free of $R_1$- and noise-bias:

$$\mathrm{PDFF} = \frac{M_f}{M_w + M_f} = \frac{\left|\frac{F_2}{C_f}\right|}{\left|\frac{W_2}{C_w} + \frac{F_2}{C_f}\right|} \qquad \text{A.5}$$

Since the terms $M_f$ and $M_w$ are used in a ratio and assuming that $B_1^-$ effects are the same for the water and fat signal components, the proposed correction for $R_1$-bias depends on $R_{1f}$ and $R_{1w}$ exclusively. Noise bias is eliminated by following a similar approach as proposed by Liu et al. [2] using the term $\left|\frac{W_2}{C_w} + \frac{F_2}{C_f}\right|$ in the denominator. Eq. A.5 is used to calculate PDFF when $\frac{|F_2|}{|W_2|+|F_2|}$<50%. When $\frac{|F_2|}{|W_2|+|F_2|}$ ≥50%, PDFF is estimated as:

$$\mathrm{PDFF} = 1 - \frac{M_f}{M_w + M_f} = 1 - \frac{\left|\frac{W_2}{C_w}\right|}{\left|\frac{W_2}{C_w} + \frac{F_2}{C_f}\right|} \qquad \text{A.6}$$

**A.2 Calculation of variance based on Cramér-Rao Lower bounds for fat and water specific R₁**

Cramér-Rao Bounds (CRB) theory enables the calculation of the minimum variance of the water- and fat-specific MP2RAGE signals. For real data or magnitude data with polarity restoration [3], the water- and fat-specific MP2RAGE signal equations becomes:

$$S_x = \frac{x_1 x_2}{x_1^2 + x_2^2} \qquad \text{A.7}$$

The variance of the MP2RAGE signal can be estimated as:

$$\sigma_{S_x}^2 = \left(\frac{\partial S_x}{\partial x_1}\right)^2 \sigma_{x_1}^2 + \left(\frac{\partial S_x}{\partial x_2}\right)^2 \sigma_{x_2}^2 + 2\frac{\partial S_x}{\partial x_1}\frac{\partial S_x}{\partial x_2}\sigma_{x_1,x_2} = \left(\frac{\partial S_x}{\partial x_1}\right)^2 \sigma_{x_1}^2 + \left(\frac{\partial S_x}{\partial x_2}\right)^2 \sigma_{x_2}^2 \qquad \text{A.8}$$

where $\sigma_{S_x}^2$ is the variance of the MP2RAGE signal, $\sigma_{x_1}^2$ and $\sigma_{x_2}^2$ are the variances of the water or fat signal magnitudes from the individual RAGE blocks, and $\sigma_{x_1,x_2}$ is the covariance between the signals from the RAGE blocks. The last term containing $\sigma_{x_1,x_2}$ is set to 0 because the signals in the RAGE1 and RAGE2 blocks are assumed to be uncorrelated, per a previously published derivation of the variance of the MP2RAGE signal equation [1].

According to CRB theory, the variance of an unbiased estimator is bounded by the corresponding diagonal component of the inverse of the Fisher Information Matrix (FIM). Thus, for the RAGE1 and RAGE2 blocks, the lower bounds of the variance for the water and fat signal components are $\sigma_{x_r}^2 \geq \sigma_{\mathrm{Signal}}^2[\mathrm{FIM}_r^{-1}]_{xx}$. $\sigma_{\mathrm{Signal}}^2$ is the variance of the signal in the RAGE blocks, which is assumed to be the same in both blocks. $[\mathrm{FIM}_r^{-1}]_{xx}$ represents the diagonal entries of the inverse of the Fisher Information matrix, $[\mathrm{FIM}_r^{-1}]_{ww}$ and $[\mathrm{FIM}_r^{-1}]_{ff}$ are the terms for the water and fat signal magnitudes, respectively. For high SNR, it can be assumed that the magnitude and real part of the signals have the same noise standard deviation [4,5]. Using this information, it is possible to get a lower bound for $\sigma_{S_x}^2$ in terms of $\sigma_{\mathrm{Signal}}^2$.

The water and fat-specific MP2RAGE signals can be assumed to depend on $R_{1x}$ exclusively, thus:

$$\sigma_{S_x}^2 = \left(\frac{dS_x}{dR_{1x}}\right)^2 \sigma_{R_{1x}}^2 \tag{A.9}$$

Then, combining Eq. A.8, A.9, and the bounds derived from CRB theory:

$$\sigma_{R_{1x}}^2 \geq \left[\frac{\left(\frac{\partial S_x}{\partial x_1}\right)^2 [FIM_1^{-1}]_{xx} + \left(\frac{\partial S_x}{\partial x_2}\right)^2 [FIM_2^{-1}]_{xx}}{\left(\frac{dS_x}{dR_{1x}}\right)^2}\right] \sigma_{Signal}^2 \tag{A.10}$$

**A.3 Calculation of variance based on Cramer-Rao Lower bounds for signal fat fraction**

Using the magnitude of the water and fat signals, the signal fat fraction (SFF) can be calculated as:

$$SFF = \frac{f_2}{w_2 + f_2} \tag{A.11}$$

The variance of this expression can be calculated as:

$$\sigma_{SFF}^2 = \left(\frac{\partial SFF}{\partial w_2}\right)^2 \sigma_{w_2}^2 + \left(\frac{\partial SFF}{\partial f_2}\right)^2 \sigma_{f_2}^2 + 2\left(\frac{\partial SFF}{\partial w_2}\right)\left(\frac{\partial SFF}{\partial f_2}\right) \sigma_{w_2,f_2} \tag{A.12}$$

where $\sigma_{w_2}^2$ and $\sigma_{f_2}^2$ are linked to the minimum theoretical variances estimated by the CRBs through $\sigma_{w_2}^2 \geq \sigma_{Signal}^2 [FIM_2^{-1}]_{ww}$ and $\sigma_{f_2}^2 \geq \sigma_{Signal}^2 [FIM_2^{-1}]_{ff}$. $\sigma_{w_2,f_2}$ is the covariance of the fat and water signal magnitudes and it can be obtained from the non-diagonal entry of the Fisher information matrix $[FIM_2^{-1}]_{fw}$.

**A.4 Calculation of variance based on Cramér-Rao Lower bounds for proton density fat fraction**

The variance of PDFF can be derived from Eq. A.5 as:

$$\sigma_{PDFF}^2 = \left(\frac{\partial PDFF}{\partial M_w}\right)^2 \sigma_{M_w}^2 + \left(\frac{\partial PDFF}{\partial M_f}\right)^2 \sigma_{M_f}^2 + 2\left(\frac{\partial PDFF}{\partial M_w}\right)\left(\frac{\partial PDFF}{\partial M_f}\right) \sigma_{M_w,M_f} \tag{A.13}$$

where $\sigma_{M_w}^2$ and $\sigma_{M_f}^2$, are the variances of $M_w$ and $M_f$, respectively. The term $\sigma_{M_w,M_f}$ is the covariance between $M_w$ and $M_f$. The variances $\sigma_{M_w}^2$ and $\sigma_{M_f}^2$ can be estimated as:

$$\sigma_{M_x}^2 = \left(\frac{1}{C_x}\right)^2 \sigma_{x_2}^2 + \left(\frac{x}{C_x^2}\right)^2 \sigma_{C_x}^2 - 2\left(\frac{x}{C_x^3}\right) \sigma_{x_2,C_x} \tag{A.14}$$

where $\sigma_{C_x}^2$ is the variance of $C_w$ or $C_f$ and $\sigma_{x_2,C_x}$ is the covariance between the fat or water signal magnitudes with their corresponding corrections.

The covariances $\sigma_{M_w,M_f}$ and $\sigma_{x_2,C_x}$ cannot easily be estimated theoretically. Consequently, for CRB-based calculations these terms were neglected. However, the correlation terms can be estimated through Monte Carlo simulations and phantom experiments as shown in Fig. C.5 and C.9.

## B Supplementary table

### B.1 Parameters for sequence optimization and numerical simulations

Table B.1 summarizes all parameter values and ranges used for sequence optimization and numerical simulations.

- Column 2 presents the parameters to calculate the MP2RAGE lookup tables (LUTs) in Fig. C.1.
- Column 3 presents the parameters for MP-optimization and the VNR/SNR heat-maps in Fig. 2. All calculations used a six-resonance fat spectrum for peanut oil retrieved from literature [6]. An infinite readout bandwidth was assumed, making chemical-shift and field-inhomogeneity-induced spatial misregistration negligible [7]. $R_{1f}$ represents substances like dairy cream [8], safflower oil [8], and fat in the human body [9]. $R_{1w}$ represents a wide range of tissues and organs [9]. $TE_1$, $\Delta TE$, and K were selected since they are reasonable choices for CSE fat-water separation and were achievable in our phantom experiments [10]. $\phi_{f_{1,2}}$ (phase of complex fat signal in RAGE 1 and 2 blocks), $\phi_{w_{1,2}}$ (phase of complex water signal in RAGE 1 and 2 blocks), $R_2^*$, eff, $\psi$ , $\phi$ , and $\varepsilon$ were selected to match values used in literature [11,12].
- Column 4 presents the parameters for ME-optimization and the VNR/SNR heat-maps in Fig. 3. $TI_1$, $TI_2$, and $TR_{MP2RAGE}$ were selected according to MP-optimization results and values achievable in phantom experiments. We considered the same criteria as in MP-optimization to select fat spectrum, $R_{1f}$, $R_{1w}$, $R_2^*$, $\phi_{f_{1,2}}$, $\phi_{w_{1,2}}$, eff, $\psi$ , $\phi$, and $\varepsilon$.
- Column 5, 6, and 7 presents the parameters for MC simulations to generate Fig. 4, C.5, and C.6, respectively. Sequence parameters $TI_1$, $TI_2$, $TR_{MP2RAGE}$, $TE_1$, and $\Delta TE$ were selected according to results from MP- and ME-optimization. Other parameters (fat spectrum, $R_{1f}$, $R_{1w}$, $R_2^*$, $\phi_{f_{1,2}}$, $\phi_{w_{1,2}}$, eff, $\psi$ , $\phi$, and $\varepsilon$) were selected following the same criteria as in MP- and ME-optimization.

**Table B.1:** Summary of parameters for sequence optimization and numerical simulations. Values considered in each case are reported as single values or ranges in brackets (discretized using a fixed step size reported in parenthesis).

| | Qualitative comparison of MP2RAGE-$T_1$ lookup tables | MP-optimization | ME-optimization | Accuracy of $R_{1f}$, $R_{1w}$, $R_2^*$, and PDFF (MC simulation) | Accuracy of estimates across PDFF and SNR (MC simulation) | Validation of VNR/SNR calculation (MC calculation) |
|---|---|---|---|---|---|---|
| **$T_1$ global [s]** | [0.05,5.00] (0.05) | - | - | - | - | |
| **PDFF** | - | [0.00,1.00] (0.25) | [0.00,1.00] (0.25) | [0.00,1.00] (0.05) | [0.00,1.00] (0.01) | [0.00,1.00] (0.07) |
| **$R_{1f}$ [$s^{-1}$]** | - | [1.20,5.00] (0.5) | 1.40, 2.00, 3.33 | [1.00,5.00] (0.16) | 3.33 | 3.33 |
| **$R_{1w}$ [$s^{-1}$]** | - | [0.286,0.986] (0.1) | 0.67, 1.00, 2.00 | [0.20,1.20] (0.04) | 0.67 | 0.67 |
| **$R_2^*$ [$s^{-1}$]** | - | [20,80] (20) | [20,80] (20) | [10,100] (3.6) | 20 | 20 |
| **$\phi_{f_{1,2}}$ [rad]** | - | $\pi/4$ | $\pi/4$ | $\pi/4$ | $\pi/4$ | $\pi/4$ |
| **$\phi_{w_{1,2}}$ [rad]** | - | $\pi/4$ | $\pi/4$ | $\pi/4$ | $\pi/4$ | $\pi/4$ |
| **SNR** | - | 30 | 30 | 30 | [5,95] (10) | 100 |
| **$TI_1$ [s]** | [0.4,1.0] (0.1) | [0.6,1.8] (0.1) | 0.6 | 0.6 | 0.6 | 0.6 |
| **$TI_2$ [s]** | [1.6,2.2] (0.1) | [1.8,3.0] (0.1) | 2.2 | 2.2 | 2.2 | 2.2 |
| **$\alpha_1$ [°]** | [1,7] (1) | 4 | 4 | 4 | 4 | 4 |
| **$\alpha_2$ [°]** | [1,7] (1) | 4 | 4 | 4 | 4 | 4 |
| **$T_{MP2RAGE}$ [s]** | [2.0,5.0] (0.5) | 4 | 4 | 4 | 4 | 4 |
| **n** | [80,140] (10) | 104 | 104 | 104 | 104 | 104 |
| **eff** | [0.94,1.06] (0.02) | 0.96 | 0.96 | 0.96 | 0.96 | 0.96 |
| **$TE_1$ [ms]** | - | 1 | [0.5,1.5] (0.1) | 1 | 1 | 1 |
| **$\Delta TE$ [ms]** | - | 1.6 | [0.1,2.0] (0.1) | 0.9 | 0.9, 1.6 | [0.0,1.0] (0.05) |
| **K** | - | 6 (unipolar) | 3 (unipolar),<br>6(unipolar),<br>6 (bipolar),<br>10 (bipolar) | 10 (bipolar) | 3 (unipolar),<br>6(unipolar),<br>6 (bipolar),<br>10 (bipolar) | 10 (bipolar) |
| **TR [ms]** | [6,10] (1) | $2TE_1+(K-1)\Delta TE$ | $2TE_1+(K-1)\Delta TE$ | $2TE_1+(K-1)\Delta TE$ | $2TE_1+(K-1)\Delta TE$ | $2TE_1+(K-1)\Delta TE$ |
| **$\psi$ [Hz]** | - | $\pi/20$ | $\pi/20$ | $\pi/20$ | $\pi/20$ | $\pi/20$ |
| **Bandwidth [Hz/pixel]** | - | Infinite | Infinite | Infinite | Infinite | Infinite |
| **$\phi$ [rad]** | - | 0 | 0, $0.02\pi$ | $0.02\pi$ | 0, $0.02\pi$ | $0.02\pi$ |
| **$\varepsilon$** | - | 0 | 0, 0.03 | 0.03 | 0, 0.03 | 0.03 |

## C. Supplementary figures

### C.1 MP2RAGE lookup table analysis for sequence optimization

$TI_1$, $TI_2$, $\alpha_1$, $\alpha_2$, and $TR_{MP2RAGE}$ have the largest impact in the shape of the MP2RAGE-$T_1$ LUTs. Conversely, TR, n, and eff have a low impact. Fig. C.1 shows the LUTs calculated when varying one parameter at a time. Fig. C.1 a) shows that TIs substantially affect the linear portion of the lookup table, especially for $T_1 \in [0.2,0.8]$ s (fat-specific $T_1$). Large TI values reduce precision and accuracy for this range of values. Fig. C.1 b) shows that the flip angles influence the inflection point at large $T_1$ values (characteristic of water-specific $T_1$) and the linear range covering $T_1 \in [0.2,0.8]$ s. Small flip angles benefit fat-specific $T_1$ mapping but also shift the inflection point limiting the measurable $T_1$ range. Fig. C.1 c) shows that $TR_{MP2RAGE}$ mostly affects the location of the inflection curve for long $T_1$ values. Long $TR_{MP2RAGE}$ values cause the inflection of the curve to happen at shorter $T_1$ values. Fig. C.1 d), e), and f) indicate that TR, n, and eff have a low impact in the shape of the MP2RAGE-$T_1$ LUTs.

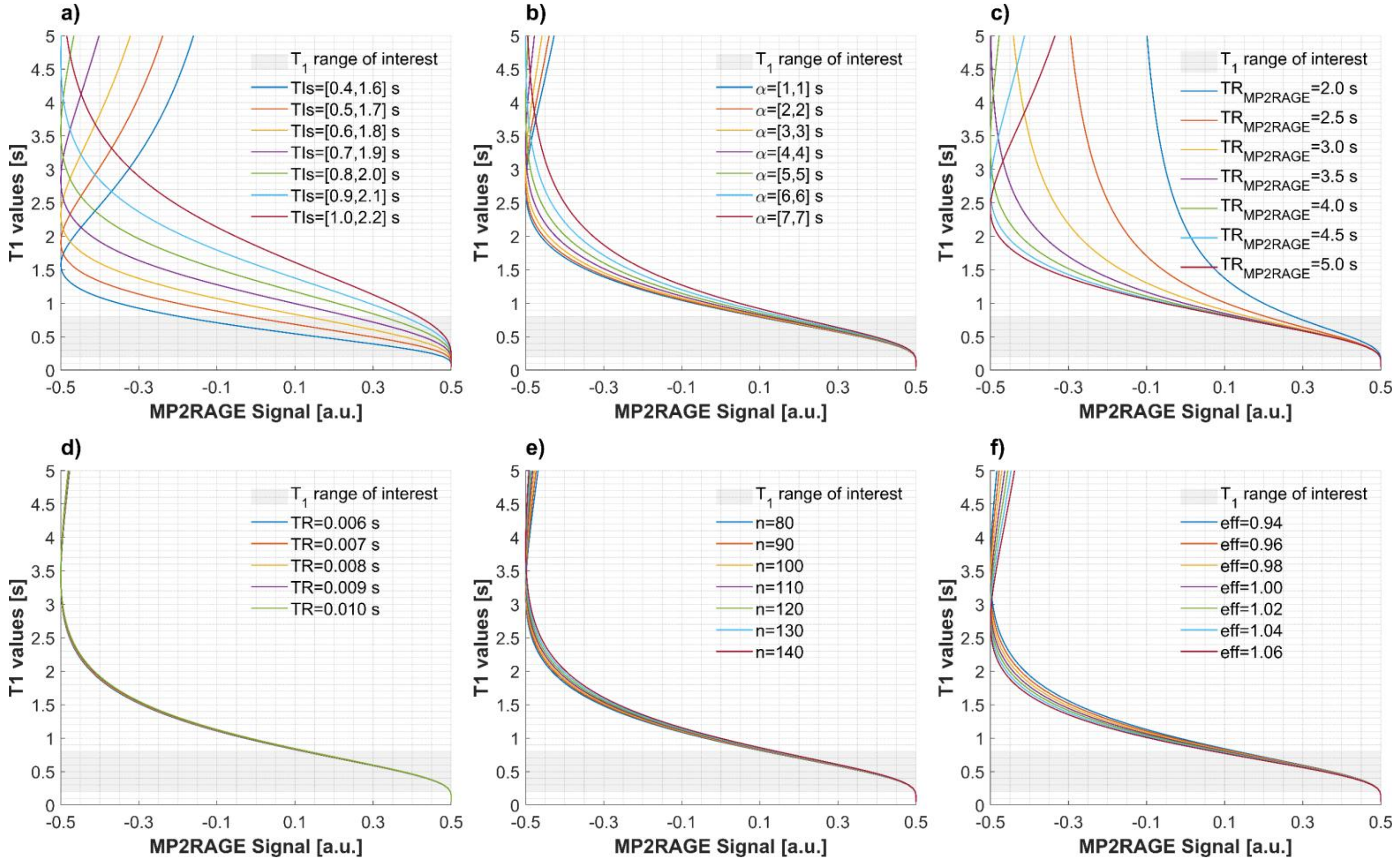

**Fig. C.1:** Qualitative comparison of the MP2RAGE-$T_1$ LUTs to illustrate effect of varying a single input while holding others constant. The varying parameters are a) $TI_1$ and $TI_2$, b) $\alpha_1$ and $\alpha_2$, c) $TR_{MP2RAGE}$, d) TR, e) n, and f) eff. The range of interest for $T_1 \in [0.2,0.8]$ s, shaded in gray, is characteristic of fatty tissue. The constant parameters are $TI_1$=0.5 s, $TI_2$=1.5 s, $\alpha_1 = \alpha_2 = 3°$, $TR_{MP2RAGE}$=2.5 s, TR=6 ms, n=100, and eff=0.96.

## C.2 Effect of transmit field inhomogeneities $B_1^*$ in the MP2RAGE lookup table

Flip angle selection for RAGE1 and RAGE2 ($\alpha_1$ and $\alpha_2$) plays a critical role in determining the sensitivity of the acquisition to $B_1^+$ inhomogeneities. Fig. C.2 shows LUT variations and the resulting $T_1$ bias for ±10% and ±20% deviations from the nominal flip angles. For $\alpha_1$=$\alpha_2$=3° (Fig. C.2 a–b), ±20% deviations yield maximum biases of 1.7% for $T_{1f}$ and 6.9% for $T_{1w}$. The maximum biases decrease to 0.8% and 3.2%, respectively, for ±10% deviations. Increasing the flip angles to $\alpha_1$=$\alpha_2$=4° (Fig. C.2 c–d) approximately doubles the maximum deviations, with biases of 3.1% and 14.8% for ±20%, and 1.5% and 6.6% for ±10%, respectively. Although a smaller flip angle reduces $B_1^+$ inhomogeneities-induced errors, they involve trade-offs, including lower SNR and a reduced measurable $T_1$ range. Selecting $\alpha_1$=$\alpha_2$=4° increases RAGE2 SNR by 30% compared to 3°, while extending the maximum measurable $T_1$ from 3.0 s to 3.4 s.

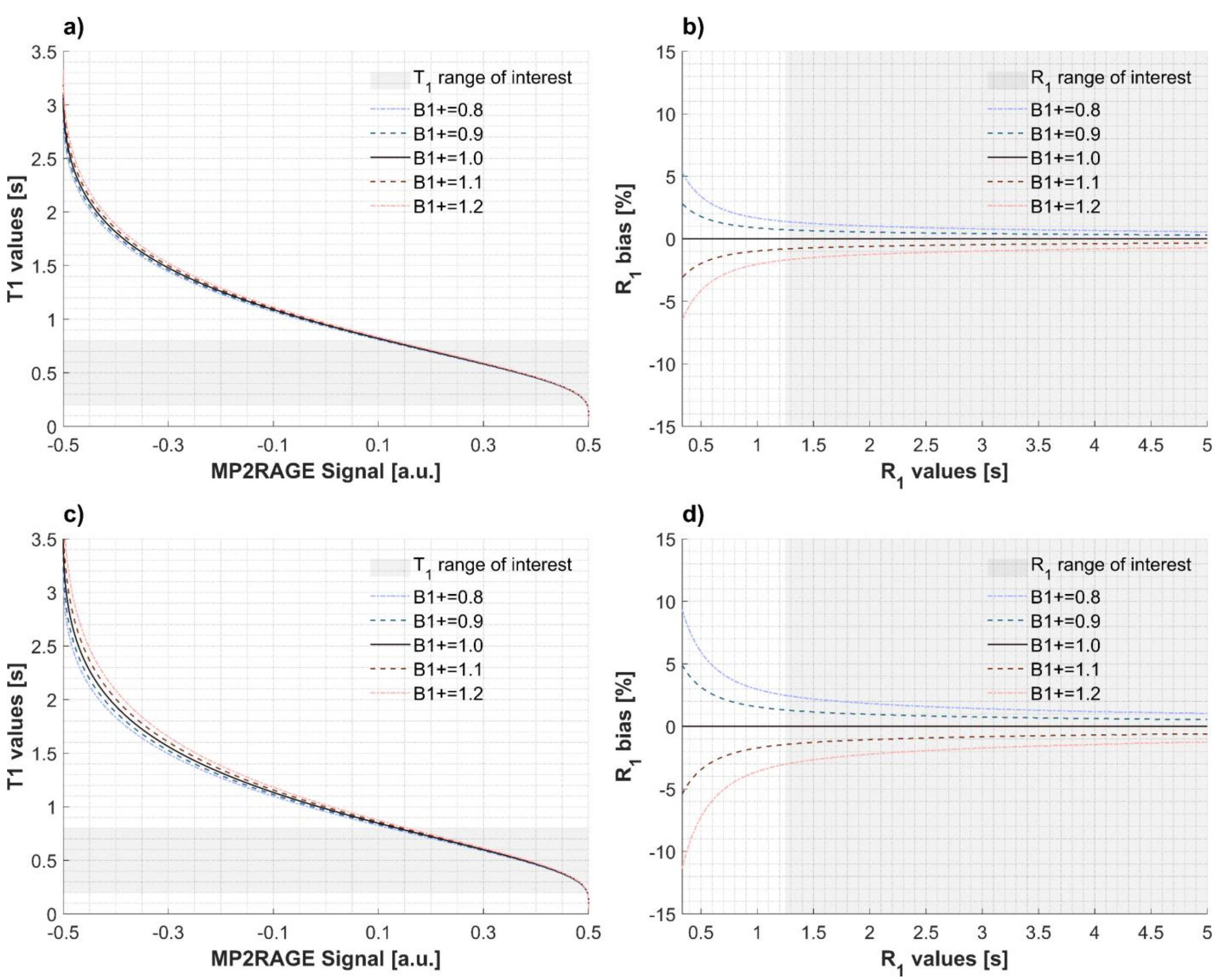

**Fig. C.2:** MP2RAGE-$T_1$ lookup tables to illustrate the effect of $B_1^+$ inhomogeneity and the resulting $T_1$ bias. a) and b) MP2RAGE-$T_1$ lookup tables and $T_1$ bias calculated for $\alpha_1$= $\alpha_2$=3° and $B_1^+$ inhomogeneity corresponding to $\pm$10% and $\pm$20% deviations. c) and d) MP2RAGE-$T_1$ lookup tables and $T_1$ bias calculated for $\alpha_1$= $\alpha_2$=4°. Calculations were performed for $TI_1$=0.6 s, $TI_2$=2.2 s, $TR_{MP2RAGE}$=4 s, TR=10 ms, n=104, and eff=0.96. These parameters match the phantom experiments. The bias was calculated as: $100\left(\frac{\text{Estimate}-\text{Ground truth}}{\text{Ground truth}}\right)$.

### C.3 Phantom diagram

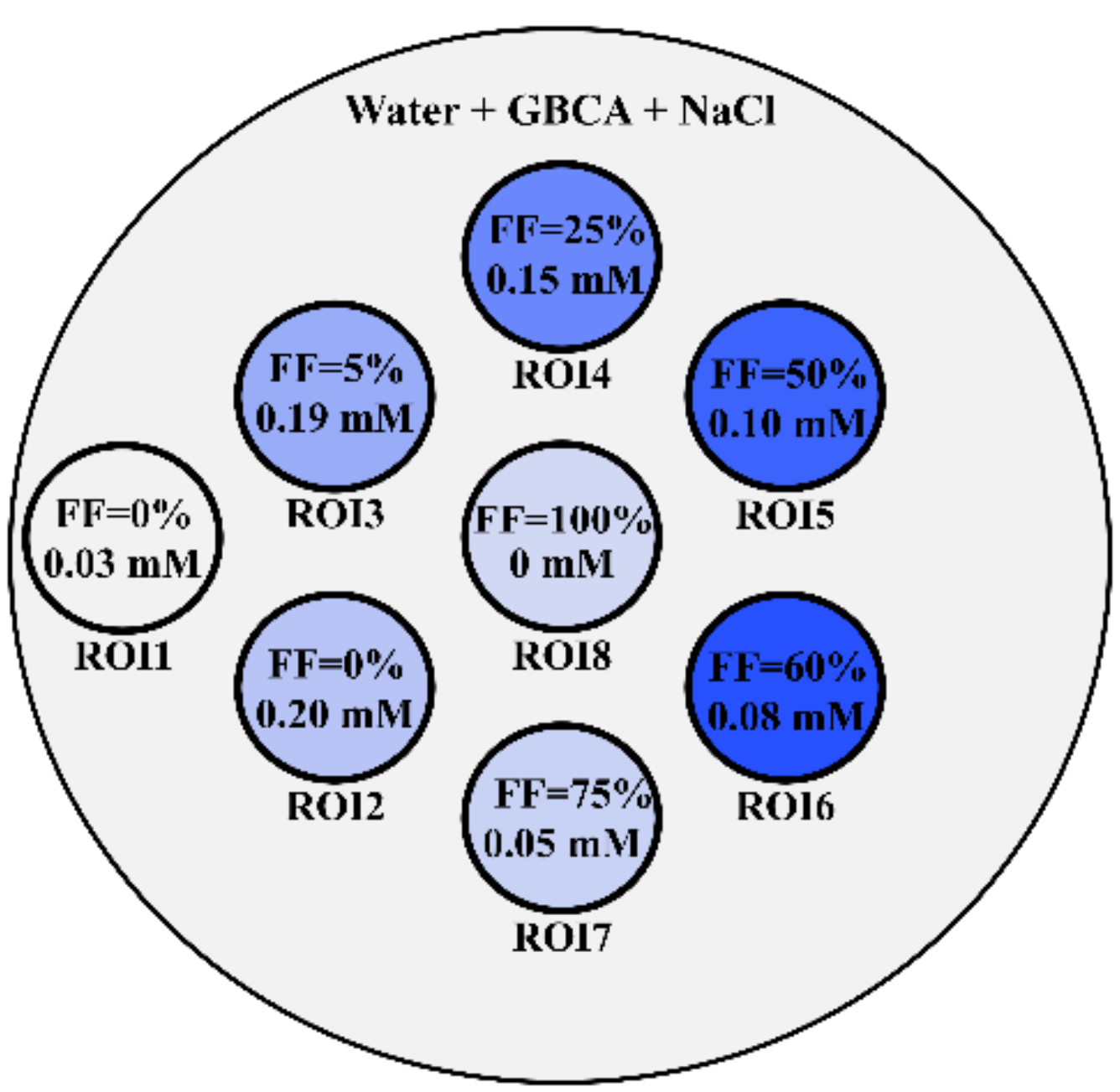

**Fig. C.3:** Phantom diagram. In the phantom, 8 ROIs are considered. ROI1 is representative of the large phantom compartment. ROIs 2-8 cover the 7 vials inside the phantom with varying fat volume fraction and GBCA concentration in an agar-oil emulsion (described in the main document). Nominal fat volume fractions and GBCA concentrations for each ROI are indicated in the diagram.

### C.4 ROIs placement for in vivo datasets

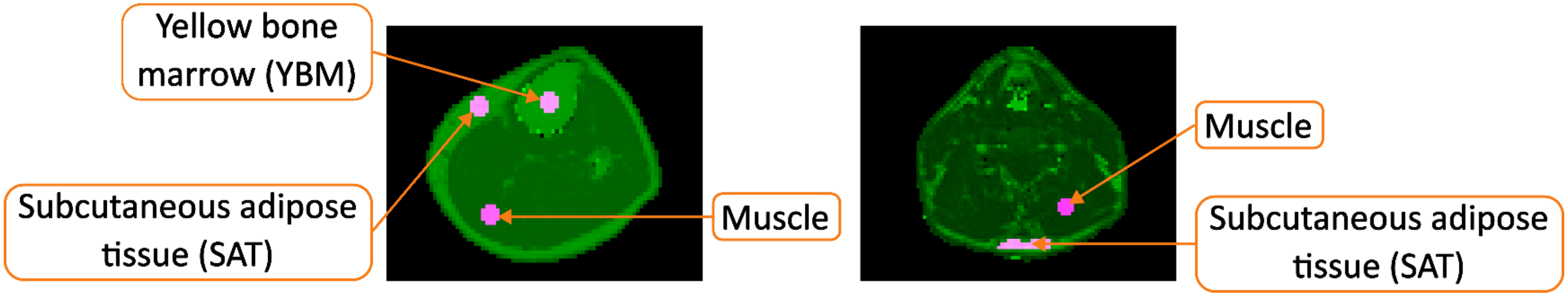

**Fig. C.4:** Anatomical ROIs for in vivo datasets. The lower leg image indicates the ROIs used for yellow bone marrow (YBM), subcutaneous adipose tissue (SAT), and muscle. The neck image indicates the ROIs for SAT, and muscle.

## C.5 Evaluation of PDFF and SNR effect on accuracy with Monte Carlo simulations

ME-MP2RAGE protocols II (6 echoes unipolar) and IV (10 echoes bipolar) are the most accurate for $R_{1f}$, $R_{1w}$, $R_2^*$, and PDFF mapping. As shown in Fig. C.5, increasing the number of echoes beyond the minimum required for ME-MP2RAGE multiparametric mapping (3 echoes unipolar or 6 echoes bipolar) substantially improves accuracy. Comparing ME-MP2RAGE II and IV to their lower-echo counterparts and for $R_{1f}$, $R_{1w}$, and $R_2^*$,: the average of the relative bias across PDFFs and SNRs decreased in 32%-98%. Similarly, for PDFF the decreased was 85%-99%.

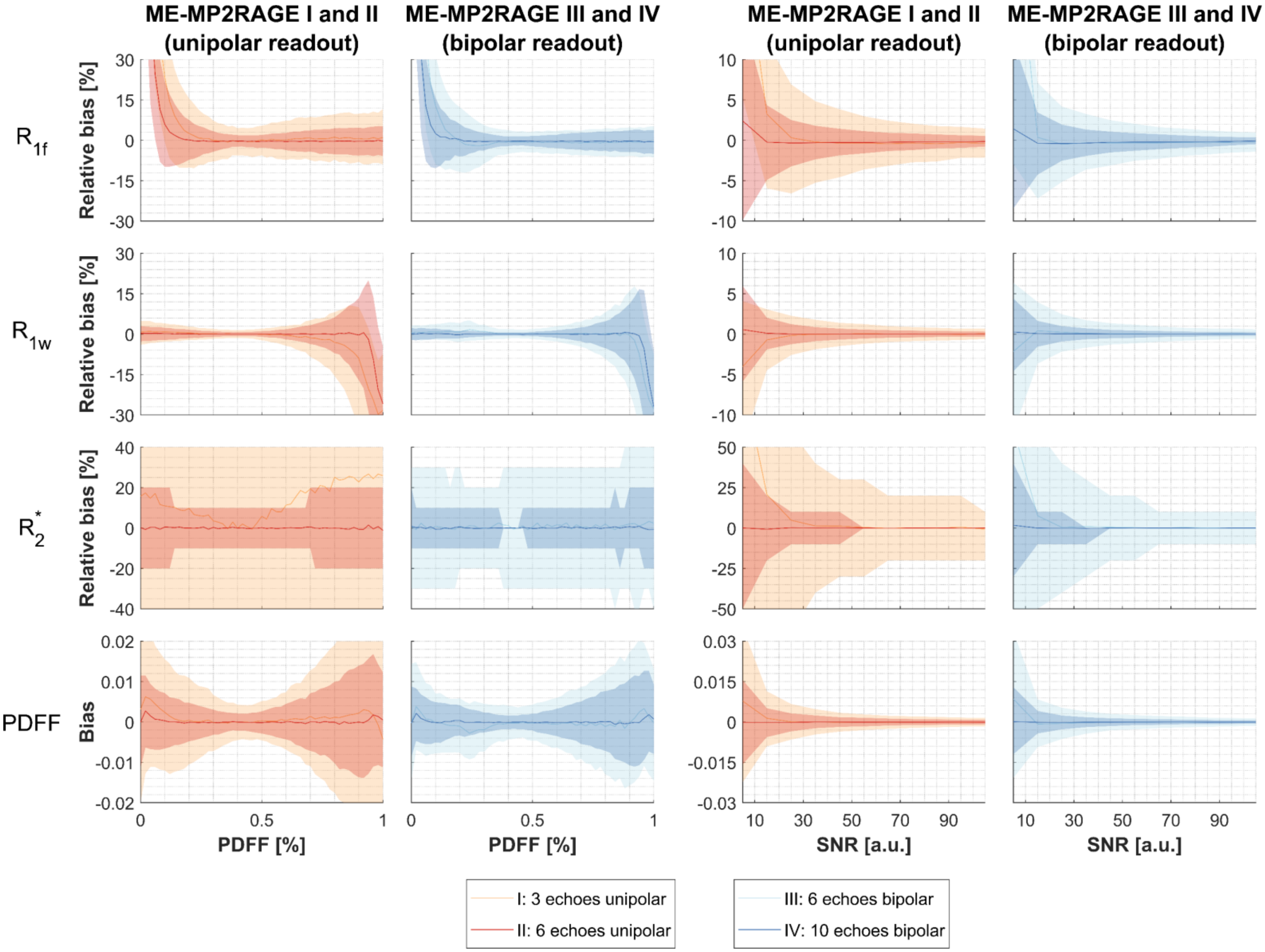

**Fig. C.5:** Evaluation of accuracy of $R_{1f}$, $R_{1w}$, $R_2^*$, and PDFF as function of PDFF and SNR. Solid line: Mean relative bias for $R_{1f}$, $R_{1w}$, $R_2^*$ and mean bias for PDFF. For PDFF and SNR analysis, SNR=30 and PDFF=0.50 were kept fixed, respectively. Shaded region: Interquartile range. Red: Results for unipolar readout gradients, ME-MP2RAGE I (3 echoes unipolar) and ME-MP2RAGE II (6 echoes unipolar). Blue: Results for bipolar readout gradients, ME-MP2RAGE III (6 echoes bipolar) and ME-MP2RAGE IV (10 echoes bipolar).

### C.6 Comparison of CRB-based calculations and MC simulations results

VNR/SNR calculated from CRB theory matches MC simulations results for $R_{1f}$, $R_{1w}$, $R_2^*$, and SFF, but not for PDFF when covariance terms are omitted (Eq. A.13 and A.14). Fig. C.6 compares VNR/SNR from CRB-based calculations and MC simulations as function of ΔTE (top row) and PDFF (bottom row). Panels a), b), c) d), f), g), h), and i) show good agreement for $R_{1f}$, $R_{1w}$, $R_2^*$, and SFF. In contrast, panels e) and j) show that neglecting covariance terms in PDFF calculations leads to CRB underestimation (orange) relative to MC results. When including covariances, results match for all parameters.

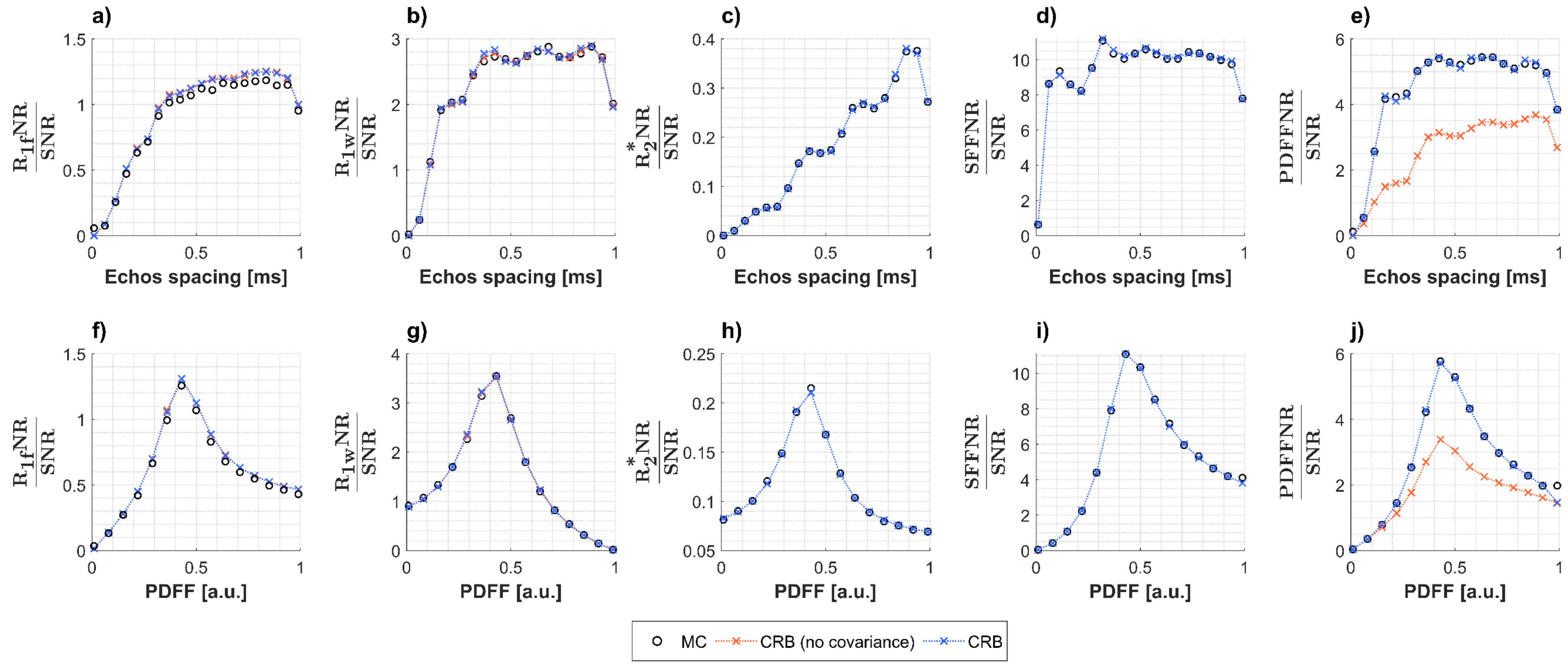

**Fig. C.6:** Comparison of CRB-based calculations and MC simulations for the VNR/SNR ratios for $R_{1f}$ (a and f), $R_{1w}$ (b and g), $R_2^*$ (c and h), SFF (d and i), and PDFF (e and j). Top row: results as function of echo spacing and fixed PDFF=0.50. Botton row: results as function of PDFF and fixed ΔTE=0.9 ms. Black markers: VNR/SNR derived from MC simulation. Orange and blue lines: VNR/SNR derived from CRB-based calculations without and with covariances, respectively. Omitting covariance terms introduces errors in PDFF calculations. MC simulations results can be used to estimate the covariances to correct the CRB calculations for PDFF.

### C.7 Multiparametric mapping with ME-MP2RAGE I, II, III, and IV

Protocols with 6 unipolar echoes (ME-MP2RAGE II) and 10 bipolar echoes (ME-MP2RAGE IV) produce better quality quantitative maps compared to protocols that used the minimum number of echoes required for multiparametric mapping with the proposed technique. Fig. C.7 presents the quantitative maps for all the ME-MP2RAGE protocols. Protocols that used the minimum number of echoes for multiparametric mapping, i.e. 3 unipolar echoes (ME-MP2RAGE I) and 6 bipolar echoes (ME-MP2RAGE III), resulted in substantially noisier quantitative maps, especially for $R_{1f}$ and $R_2^*$ maps. Poor quality of results for these two protocols justifies their exclusion from the main paper.

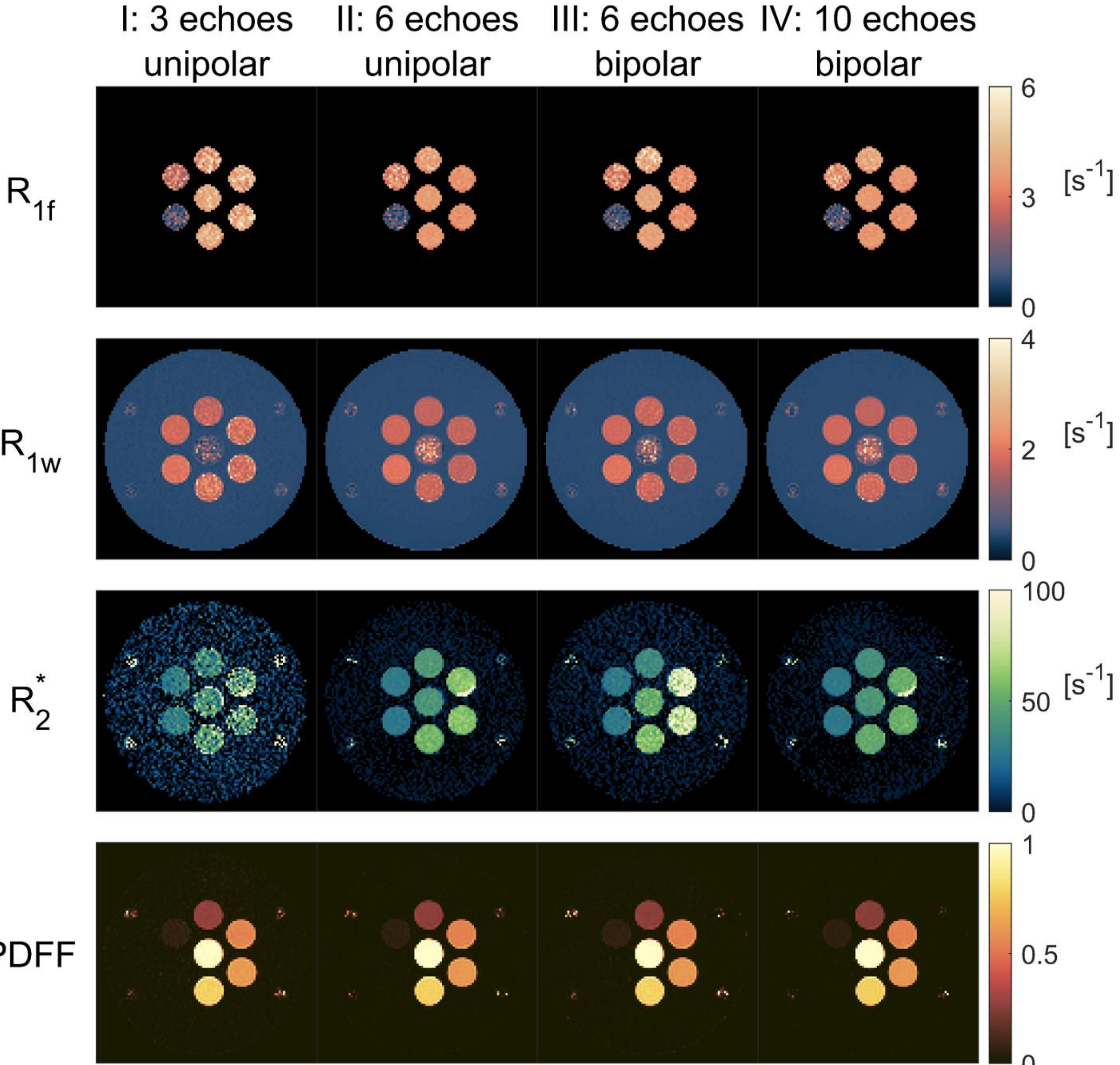

**Fig. C.7:** Quantitative maps for ME-MP2RAGE phantom experiments. Rows: From top to bottom, $R_{1f}$, $R_{1w}$, $R_2^*$, and PDFF maps. Columns: ME-MP2RAGE I-IV. $R_{1f}$ and $R_2^*$ maps for protocols I and III are substantially noisier than maps from protocols with more echoes.

### C.8 Comparison of SFF and PDFF maps in phantoms

Corrections for $R_1$- and noise-bias result in accurate PDFF estimation when using the ME-MP2RAGE sequence. Fig. C.8 compares SFF and PDFF (with corrections) maps for protocol ME-MP2RAGE IV (10 echoes bipolar). In phantoms, ROIs corresponding to nominal fat volume fractions of 25%, 50%, 60%, and 100% exhibited the largest differences between SFF and PDFF maps, whereas the 5% insert showed the smallest difference. When comparing the ME-MP2RAGE SFF and the reference 3D FLASH PDFF maps, the largest mean±95% confidence interval (CI) difference was -0.038±0.003 for the pure oil vial. The second largest difference was 0.013±0.002 for the 50% fat volume fraction emulsions. When comparing the ME-MP2RAGE PDFF map with the reference measurement, these discrepancies decreased to -0.009±0.002 and -0.005±0.002 in the pure oil and 50% inserts, respectively. Similar results were obtained for protocols ME-MP2RAGE I-III.

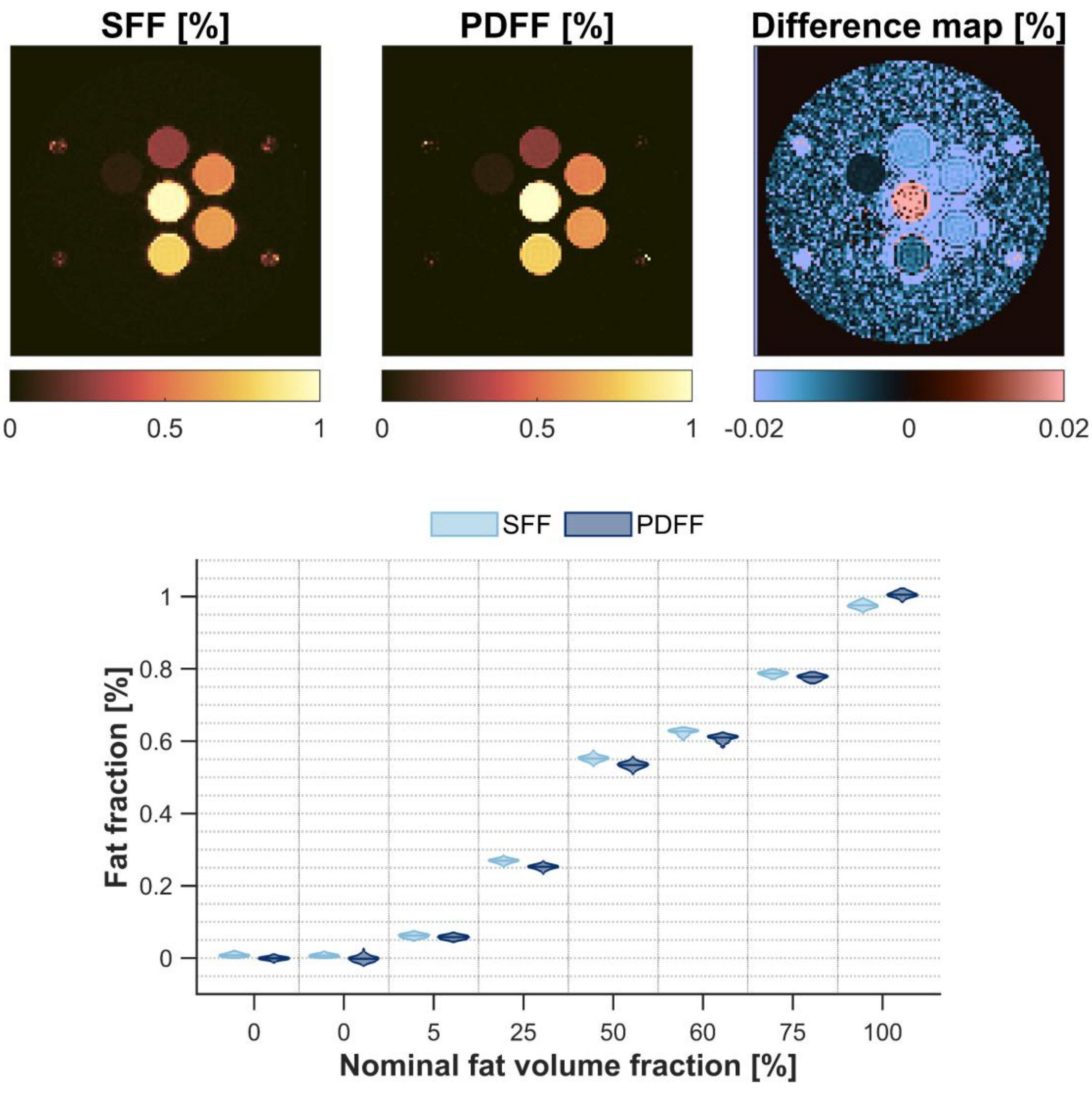

**Fig. C.8:** Comparison of SFF and PDFF maps for ME-MP2RAGE IV (10 echoes bipolar). Top maps, from left to right, show the SFF, PDFF, and difference maps (PDFF-SFF). Bottom figure shows violin plots for the SFF and PDFF measurements evaluated in the ROIs depicted in Fig. C.3. The difference map shows the regions with the largest discrepancies between the SFF and PDFF maps. The violin plot shows that, in all ROIs, the median values of PDFF maps differ from SFF maps and are closer to the nominal fat volume fraction values.

**C.9 Comparison of CRB calculations of VNR/SNR with experimental measurements in phantoms.**

CRB based calculations of VNR/SNR are representative of values measured in phantoms. Fig. C.9 shows a comparison of the median and interquartile range (IQR) of the theoretical values and values measured experimentally in phantoms for different parameters. As shown by this figure, the theoretical and experimental values follow the same trends. Consequently, optimizing sequence parameters by finding maxima of VNR/SNR is a reasonable approach since the location of these maxima (as function of sequence parameters) is expected to be similar between theoretical calculations and experiments.

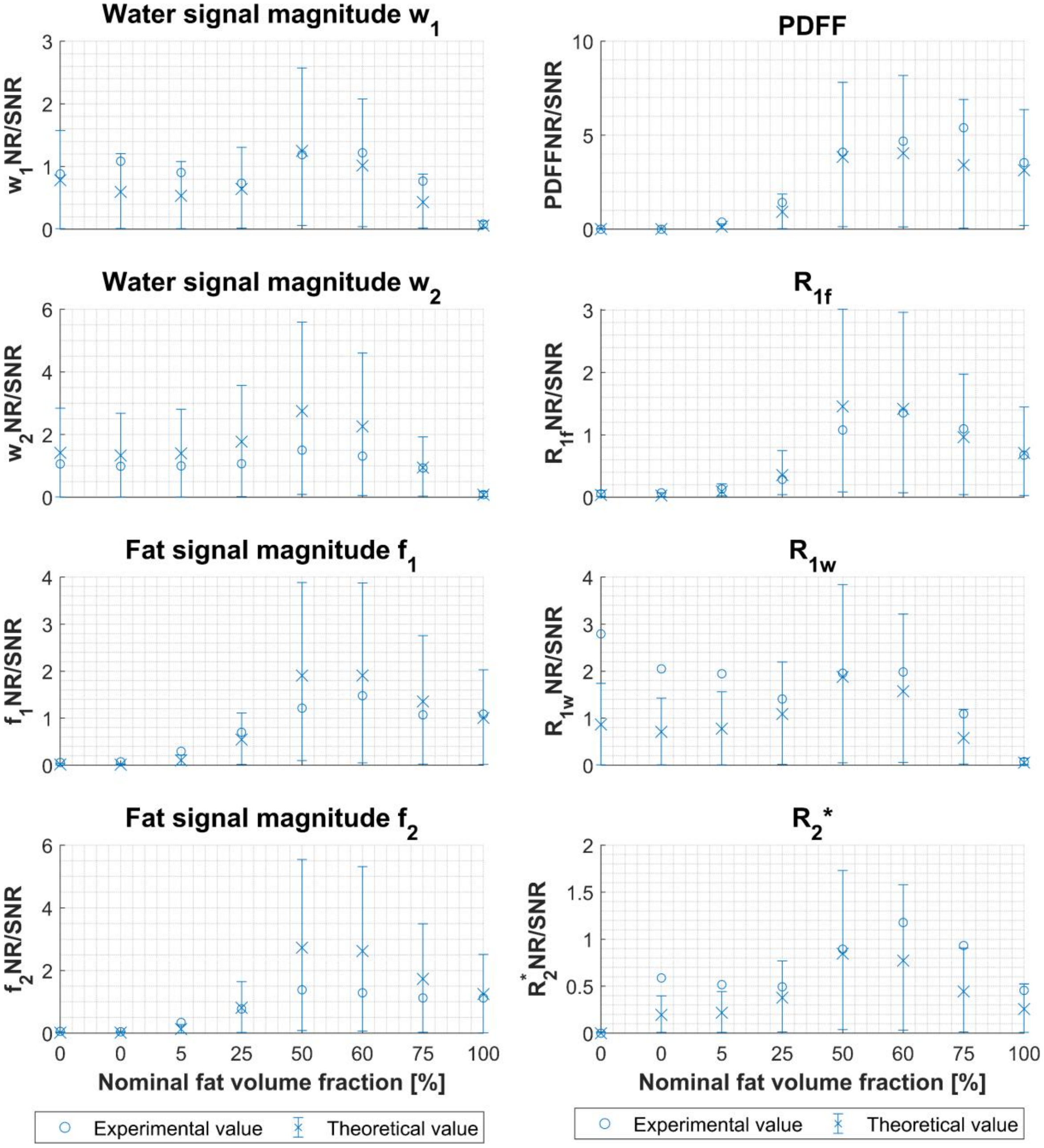

**Fig. C.9:** Comparison of experimental and theoretical VNR/SNR in phantom experiments. The experimental values correspond to the VNR values calculated as the ratio of the mean and standard deviation in each ROI of the phantom and for each parameter presented in the figure. The theoretical values are presented as the median (x marker) and the interquartile range (IQR, error bars) of the values calculated via CRB theory for each voxel of the ROIs in the phantom. In most cases, the value derived from experiments is within the IQR.

### C.10 Comparison of ME-MP2RAGE II and IV for neck datasets

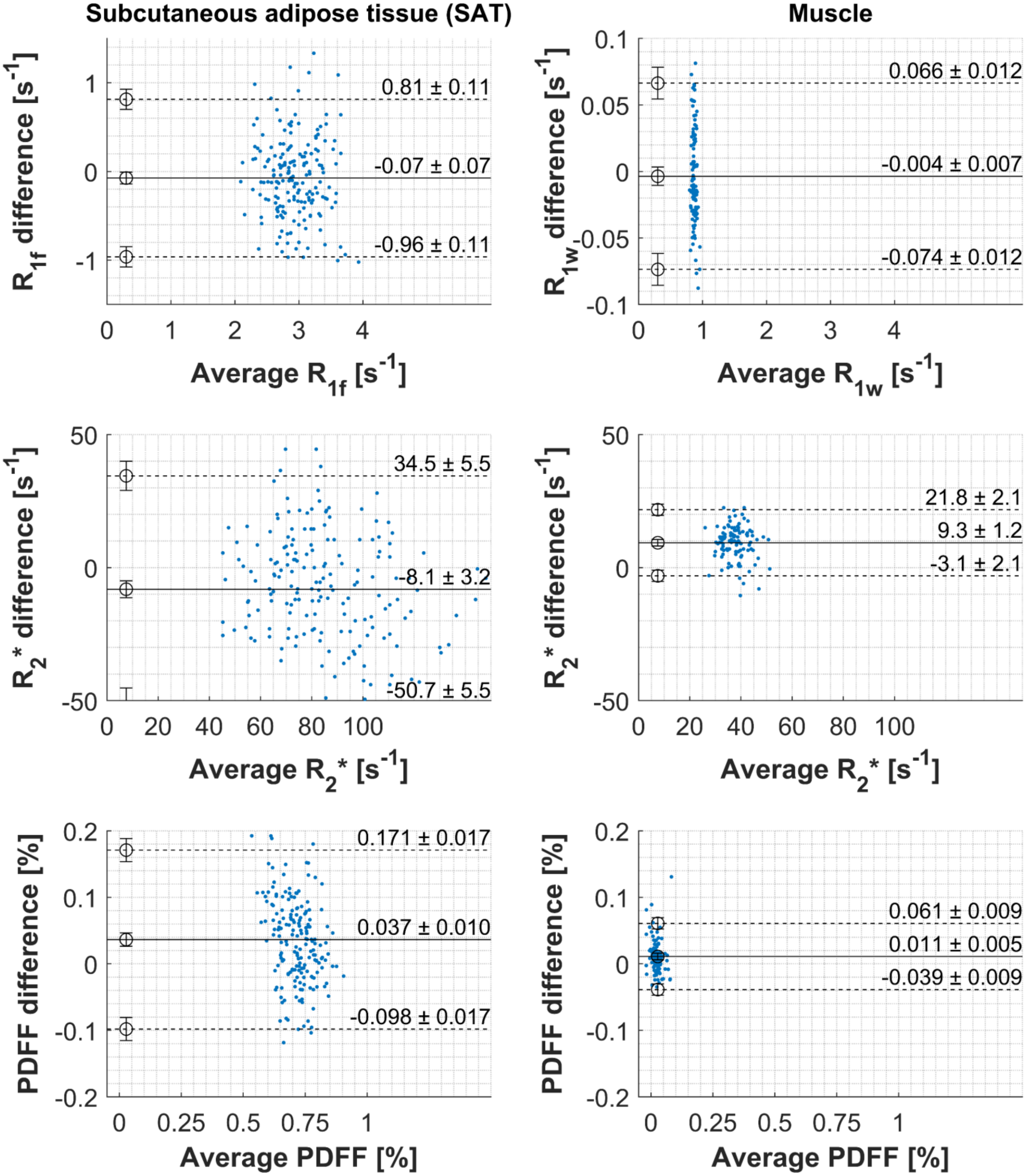

**Fig. C.10:** Bland-Altman plots comparing ME-MP2RAGE II and IV for ROIs place in the neck covering SAT, and muscle. Comparisons for $R_{1w}$ SAT, and $R_{1f}$ for muscle are omitted. Values above the solid line represent mean difference with 95% confidence intervals (CI). Values above the dashed lines represent limits of agreement with 95% CI.